\documentclass[reprint,prd,nobibnotes,nofootinbib]{revtex4-1} 
\usepackage{amsfonts}
\usepackage{graphicx}
\usepackage{amsmath}
\usepackage{mathrsfs}
\usepackage{undertilde}
\usepackage{mathtools}

\begin{document}

\title{Mass, Charge and Motion in Covariant Gravity Theories}

\author{Samuel E. Gralla}
\affiliation{Department of Physics \\ University of Maryland \\ College Park, MD 20742-411 }

\begin{abstract}
Previous work established a universal form for the equation of motion of small bodies in theories of a metric and other tensor fields that have second-order field equations following from a covariant Lagrangian in four spacetime dimensions.  Differences in the motion of the ``same'' body in two different theories are entirely accounted for by differences in the body's effective mass and charges in those different theories.  Previously the process of computing the mass and charges for a particular body was left implicit, to be determined in each particular theory as the need arises.  I now obtain explicit expressions for the mass and charges of a body as surface integrals of the fields it generates, where the integrand is constructed from the symplectic current for the theory.  This allows the entire prescription for computing the motion of a small body to be written down in a few lines, in a manner universal across bodies and theories.  For simplicity I restrict to scalar and vector fields (in addition to the metric), but there is no obstacle to treating higher-rank tensor fields.  I explicitly apply the prescription to work out specific equations for various body types in Einstein gravity, generalized Brans-Dicke theory (in both Jordan and Einstein frames), Einstein-Maxwell theory and  the Will-Nordvedt vector-tensor theory.   In the scalar-tensor case, this clarifies the origin and meaning of the ``sensitivities'' defined by Eardley and others, and provides explicit formulae for their evaluation.
\end{abstract}

\maketitle
\addtocounter{section}{1}
While general relativity (together with a cosmological constant) appears to provide a successful macroscopic description of all known gravitational phenomena, it is of interest to explore alternative theories that may provide a more fundamentally appealing description or suggest new experiments leading to the discovery of new phenomena.  A basic framework for inventing and analyzing such theories is provided by Lagrangian field theory of a Lorentz metric and other tensor fields, and indeed some version of this framework is applied in essentially all modern attempts.  In the present work we will employ this framework and restrict further to theories that are diffeomorphism covariant (no background structure), have second-order field equations, and live in four spacetime dimensions.  This class contains a majority of the theories commonly considered, and many theories that apparently violate these criteria may be rewritten so as to satisfy them.  For example, it is well known that $f(R)$ theories (with fourth-order field equations) may be rewritten as scalar-tensor theories with second-order field equations, and many higher-dimensional theories are analyzed with a four-dimensional effective Lagrangian.

Previous work \cite{gralla-motion} (hereafter paper I) considered the problem of the motion of small bodies in such theories.  Here, a ``small body'' is defined in terms of its external fields: if a region surrounding the putative body may be found such that each field approximately splits into a $1/r$-type ``body field'' plus a smoothly varying ``external field'', then one says that a small body is indeed present.  A perturbative formalism makes this notion precise, and a zero-size limit identifies the body region with a worldline in a smooth background spacetime.  We found in paper I that the equation for this worldline takes a form completely independent of both the details of the body composition and the details of the Lagrangian governing the theory.  The only parameters that appear are certain effective mass and charges of the body, which may be determined for a particular body in a particular theory by a process that involves the solution of the field equations for that body with boundary conditions representing the state of the nearby ``external universe''.  In paper I, the details of this computational process were left implicit, to be determined in each specific theory as the need arises.  The main new result of the present work is an explicit formula for the effective mass and charges that holds independent of the body and theory.

The formula is given in an abstract form in equation \eqref{hummingbird}, below, and is expressed for a scalar-vector-tensor theory in equations \eqref{Mcov}-\eqref{ehatcov}.  The main ingredient is the symplectic current $\omega^a$, which follows from a second variation of the Lagrangian, equations \eqref{varyL} and \eqref{defineomega}.  The symplectic current depends on two variations (perturbations) of the fields, and thus may be thought of as a function with two ``slots''.  Our formulae instruct one to plug in to these slots 1) the $1/r$ fields generated by the body (denoted $a/r$) and 2) a certain ``test variation'' $f$ that controls which charge one computes.  The $1/r$ fields are to be computed in the near-zone, i.e., from the far-field of a solution of the field equations for an isolated body.  One then plugs this portion of the field in to the formulae to determine the charges of that body.  In their general form these formulae involve a surface integral taken over a vanishingly small sphere centered on the position of the body in the background spacetime.  However, in any specific theory it is possible to rewrite the integral in near-zone form, i.e., over a large sphere surrounding a finite-size body.  This bypasses the need to explicitly single out the $1/r$ terms and provides a surface integral that may be applied directly on the near-zone fields.  We provide such a formula for each specific theory we consider.

A simple example that indicates the nature of our results is the classic Brans-Dicke theory \cite{fierz,jordan,brans-dicke} analyzed in the Jordan frame.  In all scalar-tensor theories of our class, the motion of a body through a background $\{g_{ab},\phi\}$ is given in terms of (non-constant) parameters $M$ (mass) and $q$ (scalar charge) by
\begin{equation}
M u^b \nabla_b u^a = q \left( g^{ab} + u^a u^b \right) \nabla_b \phi, \label{eomESintro}
\end{equation}
where $u^a$ is the four-velocity and everything is evaluated on the worldline.  To apply this formula in Brans-Dicke theory (or any specific theory), one must determine $q$ and $M$ for the body of interest.  Our formalism instructs one to take a near-zone viewpoint, where the body is of finite size and immersed in an external universe whose scale of variation is large compared to the body.  In particular, denote the value of $\phi$ on the worldline at the present time by $\hat{\phi}$.  Then, one should consider stationary solutions of the field equations with asymptotic behavior $g_{\mu \nu} \rightarrow \eta_{\mu \nu}$ and $\phi \rightarrow \hat{\phi}$, and containing a body of the type under consideration.  Having obtained such a solution, the associated charges are given by
\begin{align}
M & = \frac{1}{16\pi} \int_\infty r^2 d\Omega \ n^i \left[ \phi \left( \partial_k g_{ki} - \partial_i g_{kk} \right) - 2 \partial_i \phi  \right] \label{MBDintro} \\
q & = \frac{-1}{16\pi} \int_\infty r^2  d\Omega \ n^i \left[ 2 \frac{\omega}{\phi} \partial_i \phi + \partial_i g_{00}  + \partial_k g_{ki} - \partial_i g_{kk} \right], \label{qBDintro}
\end{align}
where $r=\sqrt{\delta_{ij}x^i x^j}$, $n^i=x^i/r$, $d\Omega$ is the area element on the unit sphere, repeated indices are summed, and $\int_\infty$ indicates an $r \rightarrow \infty$ limit of the integral.\footnote{Note that equations \eqref{MBDintro}-\eqref{qBDintro} are to be applied on the near-zone field configuration, whereas equation \eqref{eomESintro} refers to the background (far-zone) field configuration.  In the paper these different configurations are distinguished notationally and their relationship is given precisely.}  Given a stationary, asymptotically constant solution of the field equations, equations \eqref{MBDintro} and \eqref{qBDintro} provide surface-integral formulae for the mass and charge in the spirit of the Arnowitt-Deser-Misner (ADM) formula for the mass in general relativity  \cite{arnowitt-deser-misner}.  In particular, they are coordinate-invariant under the restriction of asymptotically Minkowskian metric components.  However, our mass $M$ is not derived from Hamiltonian considerations or related to time-translation symmetry in any way; instead, we simply refer to $M$ and $q$ as the ``mass'' and ``charge'' given their appearance in equation for the motion of the body, equation \eqref{eomESintro}.

To provide a deterministic prediction for the motion of a body, it is necessary not only to determine the charges $M$ and $q$ at an initial time when the body configuration is prescribed, but to have a prescription for determining $M$ and $q$ at any given time.  The simplest way of doing so is to choose enough properties of the body (such as its shape and total baryon number) such that the near-zone solution becomes a deterministic function of the asymptotic scalar field value $\hat{\phi}$.  In this case the mass and charge likewise become functions of $\hat{\phi}$ by equations \eqref{MBDintro} and \eqref{qBDintro}, and these functions may be plugged into equation \eqref{eomESintro} to determine the motion.  It is important to note, however, that not all prescriptions one might try will be consistent with the theory.  In particular, we find, with the same status as equation \eqref{eomESintro} (i.e., holding in a general theory), an equation for the evolution of the mass, 
\begin{equation}\label{massEvintro}
u^b \nabla_b M = - q u^b \nabla_b \phi.
\end{equation} 
Under circumstances where the body is assumed a deterministic function of the external field value, equation \eqref{massEvintro} implies that the functions $M(\hat{\phi})$ and $q(\hat{\phi})$ must satisfy $M'(\hat{\phi})=-q(\hat{\phi})$.  This relates our scalar charge $q$ to the sensitivity of the mass to changes in the external scalar field value, reproducing a basic feature of the classic approach of Eardley \cite{eardley}.  (Indeed, our results are completely consistent with previous work; see discussion in section \ref{sec:scalar-tensor}, below).  We go beyond previous work by treating general theories in a unified framework, by increasing the rigor of the derivation, by allowing bodies whose field configuration is not a deterministic function of the external field value, and by providing explicit surface-integral formulae for the evaluation of the mass and charge(s).

In some references the equation $q=-M'(\hat{\phi})$ is viewed as \textit{defining} the scalar charge $q$ (or equivalently the sensitivity $s=-q\hat{\phi}/M$).  This definition is less general than ours, since it requires a family of solutions (parameterized by $\hat{\phi}$), whereas our $q$ can be computed for a single solution from equation \eqref{qBDintro} (or from analogous equations in more general theories).  However, the definition is also incomplete in the sense that no instructions are given as to what property of the body is to be held fixed when the derivative with respect to $\hat{\phi}$ is taken.\footnote{It is sometimes stated that the baryon number is held fixed.  This defines the sensitivity only for bodies with an identifiable baryon number that completely specifies the body.  Examples where this definition does not work include black holes and finite-temperature stars.}  Our viewpoint turns this question around, stating that one should find a stationary solution, compute $q$ from equation \eqref{qBDintro} and use the relationship $q=-M'(\hat{\phi})$ to help \textit{determine} what properties of the body are held fixed under changes in $\hat{\phi}$.  In general, the only foolproof method of determining how a body responds to such changes is to solve the field equations for that body immersed in a slowly time-dependent external field.  However, our equation \eqref{massEvintro} indicates a particular property that \textit{any} such solution will satisfy: the mass $M$, as defined by equation \eqref{MBDintro}, will change according to the scalar charge $q$, as defined by equation \eqref{qBDintro}, by $M'(\hat{\phi})=-q$.  If one considers a body whose stationary configuration is specified by a single parameter, then the constraint $M'(\hat{\phi})=-q$ is enough to completely specify how the body changes, and therefore to completely determine its motion.  This is the case in the weak-field and black hole cases worked out in section \ref{sec:jordan-frame}, below.  For bodies specified by more than one parameter,  the constraint $M'(\hat{\phi})=-q$ only provides guidance, which must be supplemented by physical arguments or by finding slowly time-dependent solutions.

The final component of the results provided by our formalism is the equation for the perturbation to the smooth background spacetime produced by the small body.  We find for a general theory of our class that the charges appear in a universal way as delta-function sources for the linearized field equations, equations \eqref{Ephi1cov}-\eqref{Eg1cov}, below.  Written out for Brans-Dicke theory, this becomes
\begin{align}
\big[\nabla^c \nabla_c \phi\big]^{(1)} & = \frac{-8\pi}{3+2\omega} \int_\gamma\left(M+2q \phi \right) \delta_4(x,z(\tau)) d\tau \label{spork1} \\
\mathcal{E}_{ab}^{(1)} & = 8\pi \int_\gamma M u_a u_b \delta_4(x,z) d\tau, \label{spork2}
\end{align}
where superscript $(1)$ denotes linearization, the sources on the right-hand side are constructed from the background, and $\mathcal{E}_{ab}$ is the operator $\mathcal{E}_{ab}=\phi G_{ab} + \nabla^c \nabla_c \phi g_{ab} - \nabla_a \nabla_b \phi - 8 \pi T^{[\phi]}_{ab}$, with $G_{ab}$ the Einstein tensor and $T^{[\phi]}_{ab}$ the scalar-field stress-energy, given in equation \eqref{TphiBD}, below.  The integral is over the worldline of the body, denoted $\gamma$, and $\delta_4(x,x')$ is the ``covariant delta function'', given in a coordinate system by $\delta_4(x,x')=\delta(x^\mu-x'^\mu)/\sqrt{-g}$.  Once the mass $M$, charge $q$, and worldline $\gamma$ have been determined, equations \eqref{spork1} and \eqref{spork2} allow the associated waveform to be determined, making contact with far-field observables.  These equations agree with linearized versions of those obtained by Eardley  \cite{eardley}.\footnote{Eardley's equations are only formally defined, since delta-function sources appear in non-linear equations.  A natural regularization is to simply linearize his equations, and our formalism provides a rigorous justification for the resulting (well-defined) equations.}

Equations \eqref{eomESintro}-\eqref{spork2} give a simple demonstration the nature of our results using the example of Jordan frame Brans-Dicke theory.  While the new formulae \eqref{MBDintro} and \eqref{qBDintro} are certainly of interest in this classic theory, the main power of the present approach comes from its generality.  Indeed, expressions for \eqref{eomESintro}-\eqref{spork2} are given for an \textit{arbitrary} theory of our class (equations \eqref{eomcov}-\eqref{ehatcov}), and it is a simple matter to convert these expressions into useful forms for a given theory.  We have performed this procedure not only for Jordan frame Brans-Dicke theory but also for generalized scalar-tensor theories in both Jordan and Einstein frames,\footnote{One nice consequence of this analysis is the explicit demonstration of how equation \eqref{eomESintro} describes bodies in both frames, with the meaning of $q$ and $M$ changing in precisely the manner required to ensure consistency.  See equations \eqref{MES} and \eqref{qES} for the surface-integral formulae for the Einstein frame mass and charge.} for Einstein-Maxwell theory, and for the Will-Nordvedt vector-tensor theory.  In most cases we have also evaluated the charges for some known solutions and worked out the corresponding predictions of the theory.

In section \ref{sec:lagrangian} we review Lagrangian field theory and symplectic structure, establishing notational conventions.  In section \ref{sec:motion} we review the framework and results of paper I and derive the new, general formulae for the mass and charges.  In sections \ref{sec:GRCC} we apply the prescription to general relativity with a cosmological constant, while in sections \ref{sec:scalar-tensor} and \ref{sec:vector-tensor} we consider specific scalar-tensor and vector-tensor theories.

Our conventions are those of Wald \cite{wald}.  Early-alphabet lowercase latin characters $a,b,...$ are abstract spacetime indices and greek characters are spacetime coordinate indices.  When coordinates $t,x^i$ are used, mid-alphabet lowercase latin characters $i,j,...$ are spatial coordinate indices, while a $0$ denotes the time component.  We also employ a spacetime multi-index notation using mid-alphabet capital latin characters, $I,J,...$.  The meaning of these indices is described in detail at the beginning of section \ref{sec:symplectic-structure}, below.

\section{Lagrangian Field Theory}\label{sec:lagrangian}

In a series of papers spanning more than a decade \cite{wald-local, lee-wald, sudarsky-wald, wald-noether, burnett-wald, iyer-wald, iyer-wald2, gao-wald, wald-zoupas, seifert-wald}, Wald and coauthors developed a systematic and mathematically rigorous approach to Lagrangian and Hamiltonian field theory of a general $n$-dimensional covariant theory, with applications to (e.g.) black hole thermodynamics, conserved quantities, and stability.  Here we review their basic approach and present the results that are relevant for our treatment of particle motion.  In particular, we restrict their quite general results to our (still quite general) context of four spacetime dimensions and theories containing only scalar and vector fields in addition to the metric.  The extension of our methods to higher dimensions and higher rank tensor fields is left to future work.

\subsection{Lagrangian and Symplectic Structure}\label{sec:symplectic-structure}

We consider a theory of a four-dimensional Lorentz metric $g_{ab}$ along with a collection of scalar and vector fields.  However, for notational convenience, we will explicitly consider only one scalar field $\phi$ and one vector field $A^a$.  (We indicate, where necessary, how each formula would be modified for the presence of additional fields.)  We will denote the collection of all the dynamical fields $\{g,\phi,A\}$ by $\Psi$.\footnote{Note that Iyer and Wald \cite{iyer-wald} use a lowercase $\psi$ to denote the non-metric fields and a lowercase $\phi$ to denote the full set of fields.}  We often employ a multi-index notation using mid-alphabet capital latin letters,  $I,J,K,...$.  Here a lowered multi-index represents the ``natural'' position of the index on each field, $\Psi_I \sim \{g_{ab},\phi,A^a\}$, while a raised multi-index indicates the ``conjugate'' position, $\Psi^I \sim \{g^{ab},\phi,A_a\}$.  Free indices indicate the presence of an equation for each tensor field, while contracted indices indicate a sum over contractions of each tensor field, e.g., $\Psi_I \Psi^I = g_{ab} g^{ab} + \phi^2 + A^a A_a$.  When using multi-indices we do not distinguish notationally between coordinate and abstract indices.  In the coordinate index viewpoint, $\Psi_I$ stands for ``each component of each tensor field'', with the potentially confusing aspect that the multi-index position refers to the natural position, so that, e.g., the lowered-index $\Psi_I$ actually refers to the raised-index components of the vector field, $A^\mu$.  As seen below in equation \eqref{varyL}, we adopt the standard convention of varying with respect to fields in their conjugate index position (raised multi-index), so that field equation operators are defined in natural position (lowered multi-index).  Thus, we vary with respect to $\Psi^I \sim \{g^{ab},\phi,A_a\}$, producing field equation operators $E_I \sim \{E^{[g]}_{ab},E^{[\phi]},E^{[A]a}\}$.  We switch back and forth between multi-index notation and ordinary notation as necessary.  

A covariant Lagrangian field theory is specified by the choice of a scalar function $L$ that is locally and covariantly constructed \cite{wald-local} from the dynamical fields $\Psi$ and their derivatives,
\begin{equation}
L = L[g^{ab},\phi,A_a]=L[\Psi].
\end{equation}
Here and below, the use of square brackets indicates that the function considered is a local, covariant function of its arguments and their derivatives.  (We make an exception for linearized fields, as described below equation \eqref{lindef}.)  In particular, we have the covariance property,
\begin{equation}\label{Lcov}
\alpha^*L[\Psi]= L[ \alpha^*\Psi],
\end{equation}
for diffeomorphisms $\alpha$.  In the passive viewpoint, this equation states that $L$ transforms as a scalar, and we refer to $L$ as the \textit{Lagrange scalar}.  

The Lagrange scalar, as a scalar, cannot be integrated over the manifold without the specification of a volume element.  In the Wald approach, one uses the natural volume element $\epsilon_{abcd}$ associated with the metric $g_{ab}$, using a \textit{Lagrange 4-form} $L \epsilon_{abcd}$ as the fundamental object.  The Lagrange 4-form may then be integrated over the (four-dimensional) manifold $M$ without the need to specify an additional volume element.  However, for calculations done in coordinates it is useful to introduce a fixed volume element $e_{abcd}$ (also denoted $d^4x$) and define $\sqrt{-g}$ to be the scalar function such that $\sqrt{-g} e_{abcd}= \epsilon_{abcd}$.  (If $e_{abcd}$ is taken to be the coordinate volume element associated with a coordinate system, then $\sqrt{-g}$ is given by the square root of minus the determinant of the metric components in that coordinate system.)  In this case it is customary to work with a \textit{Lagrange density} $\sqrt{-g} L$, which is to be integrated with respect to the fixed volume element $e_{abcd}=d^4x$.

Many of the key results of the Wald approach to Lagrangian field theory are most simply derived and expressed in the language of differential forms \cite{iyer-wald}.  However, it can be notationally cumbersome to retain the indices associated with forms, and the results are not naturally expressed in a form that lends itself to calculation in coordinates.  While we will in some cases employ the differential forms approach in reviewing the results of the Wald approach to Lagrangian field theory, we will primarily use dualized versions of the quantities defined by Wald and collaborators (preferring scalars and vectors to 4-forms and 3-forms).  We will use the term ``Lagrangian'' to refer to the choice of $L$ in a general way, while we will be careful to distinguish between the Lagrange scalar, Lagrange 4-form, and Lagrange density when necessary.

Given a choice of Lagrangian, the action for the theory is written as (recall our notation $e_{abcd}=d^4x$),
\begin{equation}
S = \int L \sqrt{-g} \ d^4 x .
\end{equation}
The precise formulation of a variational principle based on this action requires a careful formulation of a region of spacetime and assumptions about the behavior of variations on its boundary.  Following \cite{iyer-wald}, we will avoid any associated delicate issues by defining all quantities of interest relative to the Lagrangian, without dealing with the action directly.

We now introduce the notion of a variation of the dynamical fields $\Psi$.  For a smooth one-parameter family of field configurations $\Psi(\lambda)$ (not necessarily satisfying the equations of motion), we denote the background configuration by $\Psi$ (that is, $\Psi=\Psi(\lambda=0)$) and the perturbed configuration by $\delta \Psi$ (that is, $\delta \Psi = (\partial_\lambda \Psi(\lambda))|_{\lambda=0}$).  The variation of a function $F(\Psi)$ of the fields is then defined by considering the one-parameter family $F(\Psi(\lambda))$.  Iyer and Wald \cite{iyer-wald} demonstrate that the variation of the Lagrangian may always be written
\begin{align}\label{varyL}
\delta (\sqrt{-g} L) & = \sqrt{-g} \left( E_I[\Psi] \delta \Psi^I + \nabla_a \Theta^a[\Psi,\delta \Psi] \right) ,
\end{align}
where (as explained above) the square bracket notation implies functions constructed locally and covariantly from the arguments and their derivatives. 

The tensor functions $E_I[\Psi]$ are called the field equation operators of the theory, and solutions $\Psi$ to $E_I[\Psi]=0$ are called electrovacuum solutions.  (For fields $\Psi$ that do not satisfy the field equations, we would regard $E_I[\Psi] \neq 0$ as a source; see discussion in section \ref{sec:nonev}.)   The vector function $\Theta^a$ is called the symplectic potential, defined up to an identically conserved current.  Its antisymmetrized variation is called the symplectic current $\omega^a$,
\begin{equation}\label{defineomega}
\sqrt{-g} \ \omega^a[\Psi,\delta_1 \Psi, \delta_2 \Psi] = \delta_2 (\sqrt{-g} \ \Theta^a[\Psi, \delta_1 \Psi]) - 1\leftrightarrow 2.
\end{equation}
The symplectic current may be used to construct the Hamiltonian of the theory in the manner described in \cite{iyer-wald}.  However, for our purposes we require only the following useful property,
\begin{equation}\label{divomega}
\sqrt{-g} \ \nabla_a \omega^a = \delta_1 \left( \sqrt{-g} \ E_I \right) \delta_2 \Psi^I - 1 \leftrightarrow 2.
\end{equation}
This property follows from equations \eqref{defineomega} and \eqref{varyL}, together with the fact that for a function $F$ of the dynamical fields we have $\delta_1 \delta_2 F = \delta_2 \delta_1 F$ by equality of mixed partial derivatives.   (Note that the calculation is considerably simpler if done in the language of forms, using $L \epsilon_{abcd}$ along with the hodge duals of $\Theta^a$ and $\omega^a$.  In this case one may use the fact that the exterior derivative is independent of the metric and thus commutes with variations.)  Using $\delta \sqrt{-g} = -(1/2) g_{ab} \delta g^{ab} \sqrt{-g}$, equation \eqref{divomega} may equivalently be written as
\begin{equation}
\nabla_a \omega^a = -\frac{1}{2} g_{ab} \delta_1 g^{ab} E_I \delta_2 \Psi^I + \delta_1 E_I \delta_2 \Psi^I - 1 \leftrightarrow 2.
\end{equation}
In particular, if the background configuration $\Psi$ is a solution of the field equations, then we have
\begin{equation}\label{keystone}
\nabla_a \omega^a[\Psi,\delta_1\Psi,\delta_2\Psi] = E^{(1)}_I[\delta_1 \Psi] \delta_2 \Psi^I - E^{(1)}_I[\delta_2 \Psi] \delta_1 \Psi^I,
\end{equation}
where we have defined the linearized field operators $E^{(1)}_I$ in the standard way,
\begin{equation}\label{lindef}
E^{(1)}_I[\delta \Psi] = \delta E_I.
\end{equation} 
Note that $E^{(1)}_I$ is covariantly constructed from $\Psi$ and $\delta \Psi$ (and not $\delta \Psi$ alone).  We make this one exception to our rule that the bracket notation indicates a function covariantly constructed from the displayed arguments.  Here and below, a function notated with a superscript $(1)$ will depend on a background field configuration $\Psi$ in addition to the displayed arguments.

Equation \eqref{keystone} has three important implications.  First, it shows that when $\delta_1 \Psi$ and $\delta_2 \Psi$ satisfy the linearized field equations off of an electrovacuum solution $\Psi$, then one may construct a current $\omega^a$ that is covariantly conserved on the background.  Second, it shows that the linearized field operator off of a background solution $\Psi$ is self-adjoint (according to the definition given in \cite{wald-prl}).  Third, it shows how, when the linearized field operator is acting on one field and contracted with another, one may ``move'' the operator onto the second field at the cost of a total derivative.  This type of operation occurs in the method of integration by parts, and will be of direct use to us in obtaining our surface-integral formulae for the mass and charges of a small body.

In section \ref{sec:motion} and below we will assume that the field equation operators $E_I$ are of second differential order.  In this case, the linearized field equation operators $E^{(1)}_I$ may be written
\begin{align}
E^{(1)}_I[\delta \Psi] & = \mathbf{A}_{IJ}[\Psi] \delta \Psi^J + \mathbf{B}_{IJ}^a[\Psi] \nabla_a \delta \Psi^J \nonumber \\ & \qquad + \mathbf{C}_{IJ}^{ab}[\Psi] \nabla_a \nabla_b \delta \Psi^J,
\end{align}
for some $\mathbf{A}_{IJ}$, $\mathbf{B}_{IJ}^a$, and $\mathbf{C}_{IJ}^{ab}$ covariantly constructed from the background configuration $\Psi$ and its derivatives.  Comparing with equation \eqref{keystone} and using the Leibniz rule, it follows that we may always choose the symplectic current to have the form
\begin{align}
\omega^a & = \mathbf{D}^a_{IJ}[\Psi] \delta_1 \Psi^I \delta_2 \Psi^J \nonumber \\ & \qquad + \mathbf{F}^{ab}_{IJ}[\Psi]\left( \delta_1 \Psi^I \nabla_b \delta_2 \Psi^J - \nabla_b \delta_1 \Psi^I \delta_2 \Psi^J \right), \label{omegaform}
\end{align}
for some $\mathbf{D}^a_{IJ}=\mathbf{D}^a_{JI}$ and $\mathbf{F}^{ab}_{IJ}$ covariantly constructed from the background configuration $\Psi$ and its derivatives.\footnote{One may straightforwardly attain the relationship of these quantities to $\mathbf{A}_{IJ}$, $\mathbf{B}_{IJ}^a$, and $\mathbf{C}_{IJ}^{ab}$ by direct computation.  Indeed, one may directly compute the symplectic current from the linearized equations in this manner.  However, it is usually easier to proceed from the Lagrangian via equations \eqref{varyL} and \eqref{defineomega}.}  Inherited from the corresponding freedom in $\Theta^a$, one always has the freedom to modify the symplectic current by the addition of an identically conserved current.  While we will assume in some intermediate calculations that $\omega^a$ takes the form of equation \eqref{omegaform}, the final results \eqref{Mcov}-\eqref{ehatcov} are valid for any choice of $\omega^a$.  (The invariance is explicitly ensured by Stokes' theorem.)  Equation \eqref{omegaform} holds only for theories with second-order field equations.  For higher derivative theories, an analogous form would be obtained with higher derivatives of the perturbations appearing on the right-hand side.

\subsection{Bianchi Identity}

Now consider the variation induced by a one-parameter family of diffeomorphisms (i.e., consider an infinitesimal diffeomorphism). In this case the variations of the dynamical fields are given by $\mathcal{L}_\xi \Psi^I$ for a smooth vector field $\xi$.  By the covariance property \eqref{Lcov}, the variation of the Lagrange scalar is simply $\delta L = \mathcal{L}_\xi L$, and we have
\begin{align}
\delta (\sqrt{-g} L) & = \sqrt{-g}\left( -\frac{1}{2} g_{ab} \mathcal{L}_\xi g^{ab} L + \mathcal{L}_\xi L \right) \\
& = \sqrt{-g}\left( \nabla_a \xi^a L + \xi^a \nabla_a L \right) \\
& = \sqrt{-g} \nabla_a ( \xi^a L ). \label{thing1}
\end{align}
However, from equation \eqref{varyL} we have
\begin{align}\label{thing2}
\delta (\sqrt{-g} L) & = \sqrt{-g} \left( E_I[\Psi] \mathcal{L}_\xi \Psi^I + \nabla_a \Theta^a[\Psi,\mathcal{L}_\xi \Psi] \right),
\end{align}
and combining equations \eqref{thing1} and \eqref{thing2} gives
\begin{equation}
\nabla_a \left( \Theta^a[\Psi,\mathcal{L}_\xi \Psi] - \xi^a L[\Psi] \right) = - E_I[\Psi] \mathcal{L}_\xi \Psi^I.\label{twedledee}
\end{equation}
Equation \eqref{twedledee} shows that current $\Theta^a - \xi^a L$ is conserved if $\Psi$ satisfies the field equations or if $\xi$ generates a symmetry of the dynamical fields (or both).  In the langauge of Noether's theorem, this current is simply the Noether current associated with the local symmetry\footnote{By a local symmetry we mean a variation such that $\delta (\sqrt{-g} L) = \sqrt{-g} \nabla_a \alpha^a$ for some $\alpha^a$.  Equation \eqref{thing1} shows that $\alpha^a=\xi^a L$ for infinitesimal diffeomorphisms.} of infinitesimal diffeomorphsims.  However, we may obtain a current that is conserved even when $E_I \mathcal{L}_\xi \Psi^I \neq 0$ by noting that $\mathcal{L}_\xi \Psi^I$ is linear in $\xi$ and its first derivative, so that the Leibniz rule may be used to convert this term into a term linear in $\xi$ as well as a total derivative.  To perform this operation we work with the individual fields rather than the multi-index notation.  (For the analogous calculation done in total generality, see reference \cite{seifert-wald}.)  In this case we have
\begin{align}
E_I \mathcal{L}_\xi \Psi^I & =  E^{[g]}_{ab} \mathcal{L}_\xi g^{ab} + E^{[\phi]} \mathcal{L}_\xi \phi + E^{[A]a} \mathcal{L}_\xi A_a \\
 & = E^{[g]}_{ab} (-2\nabla^a \xi^b) + E^{[\phi]} \xi^a \nabla_a \phi \nonumber \\ & \quad + E^{[A]a} ( \xi^b \nabla_b A_a + A_c \nabla_a \xi^c ) \\ 
& = \nabla^a \Big( - 2 E^{[g]}_{ab} \xi^b + E^{[A]}_{\ \ \ a} A_c \xi^c \Big) \nonumber \\
& \quad + \xi^b \Big( 2 \nabla^a E^{[g]}_{ab} + E^{[\phi]} \nabla_b \phi \nonumber \\ & \qquad \qquad + E^{[A]a} \nabla_b A_a  - \nabla_a (E^{[A]a} A_b) \Big). \label{twedledum}
\end{align}
Combining equations \eqref{twedledee} and \eqref{twedledum} gives
\begin{equation}
\nabla_a J^a_{\xi} = \xi^a B_a,
\end{equation}
with
\begin{align}
J^a_{\xi} & = \Theta^a_\xi - \xi^a L - E^{[g]}_{ab} \xi^b + E^{[A]}_{\ \ \ a} A_c \xi^c \label{Jeq} \\
B_a & = 2 \nabla^a E^{[g]}_{ab} + E^{[\phi]} \nabla_b \phi + 2 E^{[A]a} \nabla_{[b} A_{a]} - \nabla_a E^{[A]a} A_b.
\end{align}
where $\Theta^a_\xi$ reminds the reader that $\Theta^a$ is evaluated on the perturbation $\mathcal{L}_\xi \psi^I$.  Now suppose $\xi^a$ (and hence $J^a_{\xi}$) has compact support, and consider the integral (using the volume element $\epsilon=\sqrt{-g} d^4x$) of $\nabla_a J^a$ over a region containing the support region of $J^a$.  Stokes theorem implies that this integral must vanish for all $\xi^a$.  Thus the integral of $\xi^a B_a$ vanishes for all $\xi^a$, and in fact $B_a$ must vanish.  We thus derive that
\begin{align}
\nabla_a J^a_{\xi} & = 0 \label{Jcons} \\
B_a & = 0. \label{Bvan}
\end{align}
Equations \eqref{Jcons} and \eqref{Bvan} hold for arbitrary (not necessarily symmetry-generating) vector fields $\xi^a$ and arbitrary (not necessary electrovacuum) field configurations $\Psi^I$.  Thus we have an identically conserved current\footnote{Note that our $J^a_{\xi}$ differs from that of Iyer and Wald \cite{iyer-wald}, who use $J^a_\xi = \theta^a - \xi^a L$.} $J^a_{\xi}$ given by equation \eqref{Jeq} as well as a differential identity,
\begin{equation}\label{bianchi-identity}
\nabla^a E^{[g]}_{ab} - \frac{1}{2} \nabla_a E^{[A]a} A_b = E^{[A]a} \nabla_{[a} A_{b]} - \frac{1}{2} E^{[\phi]} \nabla_b \phi.
\end{equation}
We refer to equation \eqref{bianchi-identity} (equivalently equation \eqref{Bvan}) as the Bianchi identity for the theory.  We emphasize that this equation holds independent of whether the field configuration $\{g,\phi,A\}$ satisfies the field equations.  If the theory contained additional scalar or vector fields, this identity would be modified by the addition of appropriate terms of the form present above.  If the theory contained higher rank fields, there would be additional terms of similar structure; the explicit form of these terms is given in \cite{seifert-wald}.

\subsection{Non-Electrovacuum Solutions}\label{sec:nonev}

Before proceeding we make a few comments (and establish notation) regarding solutions with matter sources, or ``non-electrovacuum solutions''.  We consider a Lagrangian $L$ with action $S$ with field equation operators $E_I$, whose solutions $E_I=0$ are called electrovacuum solutions.  Suppose, however, that one wishes to consider the case where additional matter fields are present.  A common approach is to simply add an unspecified ``matter action'' $S_M$, so that the total action becomes $S' = S + S_M$.  In this approach one does not specify additional dynamical matter fields to be varied with respect to, but rather leaves $S_M$ as some unspecified function of unspecified extra fields and varies only with respect to the variables present in $S$.  In this case the equations of motion become
\begin{align}
E^{[g]}_{ab} & = \frac{-1}{\sqrt{-g}}\frac{\delta S_M}{\delta g^{ab}} \equiv \frac{1}{2} T_{ab} \label{Tfunc} \\
E^{[\phi]} & = \frac{-1}{\sqrt{-g}}\frac{\delta S_M}{\delta \phi} \equiv - \sigma \label{sigmafunc} \\
E^{[A]a} & = \frac{-1}{\sqrt{-g}}\frac{\delta S_M}{\delta A_a} \equiv J^a ,\label{Jfunc}
\end{align}
where $\delta S_M / \delta \Psi^I$ is the functional derivative, which defines the matter stress-energy $T_{ab}$, scalar charge density $\sigma$, and charge-current density $J^a$ with conventional numerical factors given above.  As a point of interpretation, note that the stress-energy of the fields $\Psi$ is not included as part of $T_{ab}$, but rather appears in the metric field operator $E^{[g]}_{ab}$ (likewise for scalar charge and charge-current).  For example, in Einstein-Maxwell theory we have $E^{[g]}_{ab} \propto G_{ab} - 8 \pi T^{EM}_{ab}$, where $T^{EM}_{ab}$ is the usual electromagnetic stress-energy tensor, so that $T_{ab}$ has the interpretation of the non-electromagnetic stress-energy.  This is the origin of our name ``electrovacuum solution'' for solutions of the field equations with vanishing $\{T_{ab},\sigma,J^a\}$.

Since not every classically admissible matter source will follow from a Lagrangian---and since the details of any matter model are irrelevant to our main results---we find it more natural to work at the level of equations of motion for the matter, rather than proceeding from a Lagrangian.  For our purposes, a ``matter model'' is a set of equations for some auxiliary matter variables (density, pressure, fluid four-velocity, etc.) together with a prescription for constructing $\{T_{ab},\sigma,J^a\}$ from these variables.  We then couple that matter model to our theory by solving
\begin{align}
E^{[g]}_{ab} & = \frac{1}{2} T_{ab} \label{nonevT} \\
E^{[\phi]} & = - \sigma \label{nonevphi} \\
E^{[A]a} & = J^a \label{nonevA}
\end{align}
together with the equations for the auxiliary matter variables.  While we do not require our sources $\{T_{ab},\rho,J^a\}$ to follow from a Lagrangian, the reader more familiar with that approach may imagine that $\{T_{ab},\rho,J^a\}$ are obtained from a matter action from functional derivatives via equations \eqref{Tfunc}-\eqref{Jfunc}.  Note that the Bianchi identity, equation \eqref{bianchi-identity}, implies that solutions will only exist when
\begin{equation}\label{bianchi-identity-nonev}
\nabla^a T_{ab} - \nabla_a J^a A_b = 2 J^a \nabla_{[a} A_{b]} + \sigma \nabla_b \phi.
\end{equation}

\section{Motion and Charges}\label{sec:motion}

We now review the methods and results of paper I, paying special attention to the appearance of the charges that characterize the small body and its motion.  In the process, we will derive the new formulae for these charges.  As in the derivation of the Bianchi identity, above, we restrict for simplicity to a theory of a metric, scalar, and vector field.  While most of the analysis of the section generalizes immediately to a general theory, in some places the assumption of a single scalar and vector field plays an important role in the form of the expressions.  In these cases we indicate how the derivation and results would change in the presence of additional fields.

\subsection{Framework and Assumptions}\label{sec:framework}

The basic philosophy of paper I (derived from that of \cite{gralla-wald}) is as follows.  We imagine a field configuration $\Psi$ (the ``physical configuration'') which contains a finite-sized body that is small compared to the scale of variation of the external universe.  (In particular, all scales associated with the body, such as mass, charge, etc., must be small compared to the scales of the external fields.)  We then seek an approximate description of the physical configuration by attaching it to a one-parameter family $\Psi(\lambda)$ to be Taylor expanded in $\lambda$.\footnote{For smooth one-parameter families, this viewpoint is equivalent to the more heuristic approach of splitting the physical field configuration into a background and a perturbation and expanding in the perturbation.  A treatment of small bodies will usually entail singularities on a worldline, in which case the equivalence is less evident and we find the formal mathematical approach significantly clearer.}  In order that the truncated Taylor expansion accurately approximate the physical configuration $\Psi$, we choose the family such that, as $\lambda \rightarrow 0$, all scales associated with the body approach zero.  Since Taylor's theorem guarantees an asymptotic approximation as $\lambda \rightarrow 0$, our perturbation series in $\lambda$ should accurately approximate the physical configuration as long as the scales of the body are small compared to the scales of the external universe.

Of course, to implement the above procedure one must identify suitable assumptions on the one-parameter family that reflect the satisfaction of the above physical criteria.  Our approach, described in detail in \cite{gralla-wald,gralla-motion}, involves the requirement that certain ``near-zone'' and ``far-zone'' limits exist and are related to each other in a smooth way.  The main output of these assumptions is an assumed form of the field perturbations in coordinates $t,x^i$ that have the interpretation of being centered on the body (i.e., the worldline $x^i=0$, called $\gamma$, describes the lowest-order position of the body).  We also introduce the distance coordinate $r=\sqrt{\delta_{ij}x^ix^j}$ as well as $n^i=x^i/r$, denoted by $\vec{n}$ when representing a direction on the sphere.  We use superscript $(m)$ to denote an $m$th order perturbation,
\begin{align}
\Psi_I(\lambda) & = \Psi_I^{(0)} + \lambda \Psi_I^{(1)} + O(\lambda^2) \label{Psidown} \\
\Psi^I(\lambda) & = \Psi^{(0)I} + \lambda \Psi^{(1)I} + O(\lambda^2), \label{Psiup}
\end{align}
where, as is evident from these equations, \textit{we do not use the unperturbed metric to raise and lower indices of the perturbations}.  This convention harmonizes more naturally with the variational viewpoint, in terms of which the symplectic current---and hence our effective mass and charges---are naturally expressed.   However, it will be useful to have a map that relates $\Psi^{(1)I}$ to $\Psi^{(1)}_I$, and we define $q_{IJ}$ and $q^{IJ}$ by 
\begin{align}
\Psi^{(1)}_{I} & = q_{IJ} \Psi^{(1)J} \label{lowerPsi} \\
\Psi^{(1)I} & = q^{IJ} \Psi^{(1)}_{J}.  \label{raisePsi}
\end{align}
In matrix language equation \eqref{lowerPsi} reads
\begin{equation}
\left( \begin{array}{c}
g^{(1)}_{ab}  \\
\phi^{(1)} \\
A^{(1)c} \end{array} \right)
=
\left( \begin{array}{ccc}
-g^{(0)}_{aa'} g^{(0)}_{bb'} & 0 & 0 \\
0 & 1 & 0 \\
A^{(0)}_{b'} \delta^{c}_{\ a'} & 0 & g^{(0)cc'} \end{array} \right)
\left( \begin{array}{c}
g^{(1)a'b'}  \\
\phi^{(1)} \\
A^{(1)}_{c'} \end{array} \right), \label{lowerPsi2}
\end{equation}
while equation \eqref{raisePsi} reads
\begin{equation}
\left( \begin{array}{c}
g^{(1)ab}  \\
\phi^{(1)} \\
A^{(1)}_{c} \end{array} \right)
=
\left( \begin{array}{ccc}
-g^{(0)aa'} g^{(0)bb'} & 0 & 0 \\
0 & 1 & 0 \\
A^{(0)b'} \delta_{c}^{\ a'} & 0 & g^{(0)}_{cc'} \end{array} \right)
\left( \begin{array}{c}
g^{(1)}_{a'b'}  \\
\phi^{(1)} \\
A^{(1)c'} \end{array} \right).\label{raisePsi2}
\end{equation}
The ``components'' of $q_{IJ}$ and its inverse $q^{IJ}$ may be read off of the matrices in equations \eqref{lowerPsi2} and \eqref{raisePsi2}, respectively.  Note that this raising and lowering convention does not allow for the raising and lowering of a single abstract index on the metric perturbation $g^{(1)}_{ab}$, but only for the raising of both indices at once.  We will use this raising and lowering convention for the majority of the paper.  However, when displaying expressions for the symplectic current, we will considerably shorten the length by raising and lowering individual indices with the background metric (and abandoning the convention described above).  In order to avoid confusion, in all such cases we will use a special symbol for the perturbation, such as $h_{\mu \nu}$ instead of $g^{(1)}_{\mu \nu}$, and in all such cases we will explicitly note the convention.  Note also that, as evidenced in equations \eqref{Psiup}-\eqref{raisePsi2}, we do not bother to maintain a constant index position with respect to perturbative superscripts (or other non-index modifiers); we happily write both $\Psi^{(0)}_I$ and $\Psi^{(0)I}$.

At finite $\lambda$, we impose the field equations only for $r > \lambda \bar{R}$ (for some $\bar{R}$), leaving the interior of this worldtube arbitrary; this ensures that our results hold independent of the internal composition of the body.  The perturbation series, then, is only defined for $r>0$.  However, we assume that the background fields are smooth (the body ``smoothly disappears''), so that we may include the curve $r=0$ in the background field configuration.  We mark the value taken at $r=0$ with a hat, writing
\begin{equation}\label{hatter}
\Psi_I(t,x^i) = \hat{\Psi}_I(t) + O(r).
\end{equation}
Here and below the $O(r^n)$ symbols are at fixed $t,\vec{n}$.  We will always set the metric on the worldline $\gamma$ to $\eta_{\mu \nu}$ by coordinate choice, so that equation \eqref{hatter} reads
\begin{align}
g^{(0)}_{\mu \nu} & = \eta_{\mu \nu} + O(r) \label{g0} \\
\phi^{(0)} & = \hat{\phi}(t) + O(r) \label{phi0} \\
A^{(0)\mu} & = \hat{A}^{\mu}(t) + O(r). \label{A0}
\end{align}
Our assumptions imply that the first perturbations take the form
\begin{align}
\Psi^{(1)}_I & = \frac{a_I(t,\vec{n})}{r} + O(r^0) & (r>0)\ \\
\Psi^{(1)I} & = \frac{a^I(t,\vec{n})}{r} + O(r^0) & (r>0),
\end{align}
i.e., the perturbations diverge as $1/r$ near $\gamma$.  We have introduced the field $a_I$ to characterize the time and angular dependence of the divergence.  As is evident for the above equations, we raise and lower indices on $a_I$ with the map $q_{IJ}$ rather than the unperturbed metric.  For the individual fields, we denote $a_I=\{a^{[g]}_{\mu \nu},a^{[\phi]},a^{[A]\mu}\}$, so that we have
\begin{align}
g^{(1)}_{\mu \nu} & = \frac{a^{[g]}_{\mu \nu}(t,\vec{n})}{r} + O(r^0) & (r>0)\ \label{g1far} \\
\phi^{(1)} & = \frac{a^{[\phi]}(t,\vec{n})}{r} + O(r^0) & (r>0)\ \label{phi1far} \\
A^{(1)\mu} & = \frac{a^{[A]\mu}(t,\vec{n})}{r} + O(r^0) & (r>0).\label{A1far}
\end{align}
Equations \eqref{Psiup}-\eqref{A1far} describe the far-zone background $\Psi^{(0)}$ and perturbations $\Psi^{(1)}$.  In this far-zone limit, the body shrinks to zero size/mass/charge and disappears entirely from the background field configuration, leaving behind the smooth ``external universe'' $\Psi^{(0)}$.  However, the body leaves behind the preferred worldline $x^i=0$ as the place where it ``disappeared to'', and the field perturbations become singular on this worldline.

We also consider an alternative, near-zone limit, which is designed to keep the body at fixed size.  While the far-zone limit represents the viewpoint of an observer far from the body, where its effects are small, the near-zone limit represents the viewpoint of an observer in the vicinity of the body, where the effects of the external universe are small.  To define this limit we first introduce rescaled coordinates $\bar{x}^i = \lambda^{-1} x^i$ and (for some time $t_0$) $\bar{t}=\lambda^{-1}(t-t_0)$.  An observer at fixed scaled coordinate moves correspondingly closer to the body as it shrinks in size, with all measured scales reducing to zero.  However, if we ``change units'' by rescaling the fields,
\begin{align}
\bar{g}_{\mu \nu} & = \lambda^{-2} g_{\mu \nu} \label{gbar} \\
\bar{\phi} & = \phi \label{phibar} \\
\bar{A}^{\mu} & = \lambda A^{\mu}, \label{Abar}
\end{align}
then the scales associated with the body (which shrank to zero in the far-zone limit) now remain fixed, and a finite $\lambda \rightarrow 0$ limit is achieved.\footnote{Note that defining rescaled fields with conjugate index position would require different powers of $\lambda$.  For simplicity we will not define such fields in this paper, working only with scaled fields in the ``natural'' index position of equations \eqref{gbar}-\eqref{Abar}.}  Note that the powers of $\lambda$ in the definitions of the rescaled fields precisely cancel the powers of $\lambda$ that arise from a coordinate transformation to scaled coordinates.  Thus the components of the scaled version of a field can always be computed by simply plugging in for the new coordinates,
\begin{align}
\bar{g}_{\bar{\mu} \bar{\nu}}(\lambda;t_0;\bar{t},\bar{x}^i) & = g_{\mu \nu}(\lambda; t=t_0+\lambda \bar{t},x^i=\lambda \bar{x}^i) \label{gplugin} \\
\bar{\phi}(\lambda;t_0;\bar{t},\bar{x}^i) & = \phi(\lambda; t=t_0+\lambda \bar{t},x^i=\lambda \bar{x}^i) \label{phiplugin} \\
\bar{A}^{\bar{\mu}}(\lambda;t_0;\bar{t},\bar{x}^i) & = A^\mu(\lambda; t=t_0+\lambda \bar{t},x^i=\lambda \bar{x}^i) .\label{Aplugin} 
\end{align}
We denote scaled coordinate components by placing a bar on the coordinate index, so that the above equations relate components in scaled coordinates of scaled fields to corresponding components of the original fields in original coordinates.  

Having introduced the scaled coordinates and fields, the scaled limit is simply the $\lambda \rightarrow 0$ limit,
\begin{align}
\bar{g}^{(0)}_{\bar{\mu}\bar{\nu}}(t_0;\bar{t},\bar{x}^i) & = \lim_{\lambda \rightarrow 0} \bar{g}_{\bar{\mu} \bar{\nu}}(\lambda;t_0;\bar{t},\bar{x}^i) \\
\bar{\phi}^{(0)}(t_0;\bar{t},\bar{x}^i) & = \lim_{\lambda \rightarrow 0} \bar{\phi}(\lambda;t_0;\bar{t},\bar{x}^i)  \\
\bar{A}^{(0)\bar{\mu}}(t_0;\bar{t},\bar{x}^i) & = \lim_{\lambda \rightarrow 0} \bar{A}^{\bar{\mu}}(\lambda;t_0;\bar{t},\bar{x}^i) , \label{cowgirl}
\end{align}
where the limit is taken at fixed scaled coordinate.  Since we work only at $r > \lambda \bar{R}$ (for some $\bar{R}$), the limiting fields $\bar{\Psi}^{(0)}$ are defined only for $\bar{r} > \bar{R}$, i.e., only for sufficiently large $\bar{r}$.  Thus, while we apply the term ``near-zone'' to this limit (and refer to $\bar{\Psi}^{(0)}$ as the ``near-zone background field configuration''), it should be borne in mind that these fields correspond to a near-zone description only of the body's exterior.  (Indeed, for this reason these fields were referred to as the ``body exterior fields'' in \cite{gralla-motion}.)  Determining the relevant near-zone fields in a given application requires matching to an interior solution, as discussed below equation \eqref{Ebar0b}, below.

An immediate consequence of equations \eqref{gplugin}-\eqref{Aplugin} is that the near-zone background fields are stationary (independent of $\bar{t}$), and, coupled with our more general assumptions, a further consequence is that the fields admit the large-$\bar{r}$ expansions,
\begin{align}
\bar{g}^{(0)}_{\bar{\mu}\bar{\nu}}(t_0;\bar{x}^i) & = \eta_{\mu \nu} + \frac{a^{[g]}_{\mu \nu}(t_0,\vec{n})}{\bar{r}} + O\left( \frac{1}{\bar{r}^2} \right) \label{gbar0} \\
\bar{\phi}^{(0)}(t_0;\bar{x}^i) & = \hat{\phi}(t_0) + \frac{a^{[\phi]}(t_0,\vec{n})}{\bar{r}} + O\left( \frac{1}{\bar{r}^2} \right) \label{phibar0} \\
\bar{A}^{(0)\bar{\mu}}(t_0;\bar{x}^i) & = \hat{A}^\mu(t_0) + \frac{a^{[A]\mu}(t_0,\vec{n})}{\bar{r}} + O\left( \frac{1}{\bar{r}^2} \right).\label{Abar0}
\end{align}
Note the dual role played by the hatted quantities as the value of the far-zone background metric as the position of the particle and the asymptotic value of the near-zone field configuration.  Similarly, note the dual role played by the $a^{[\Psi]}$ as the $1/r$ divergence of the far-zone perturbation at time $t$ and the $1/\bar{r}$ correction to the asymptotic value of the stationary near-zone field configuration computed at time $t=t_0$.

\subsubsection{The Near-zone Background Field Equations and a Subtlety for Non-Scale-Invariant Theories}\label{sec:nonscaleinv}

An important point regarding the near-zone background fields $\bar{\Psi}^{(0)}$ is that that these fields do not in general satisfy the full field equations, but rather obey those of a ``scale-invariant sub-theory''.  For the scaled variables at finite $\lambda$, we define
\begin{equation}\label{Ebar}
\bar{E}^{[\Psi]}[\bar{g},\bar{\phi},\bar{A}] \equiv E^{[\Psi]}[\lambda^2 \bar{g},\bar{\phi},\lambda^{-1}\bar{A}]
\end{equation}
for each field $\Psi$.  This defines an operator on field configurations $\bar{E}^{[\Psi]}$ as a function of $\lambda$.  If the equations scale homogeneously, $E^{[\Psi]}[\lambda^2 g,\phi,\lambda^{-1} A]=\lambda^N E^{[\Psi]}[g,\phi,A]$ for some integer $N$, then the operator agrees (up to powers of $\lambda$) with the original operator $E^{[\Psi]}$, and the scaled fields satisfy the same equations as the original fields.  If not, then in general the scaled fields will satisfy a different equation.  In particular the relevant operator for the background scaled fields is
\begin{equation}\label{Ebar0a}
\bar{E}^{[\Psi](0)}[\bar{g},\bar{\phi},\bar{A}] \equiv \lim_{\lambda\rightarrow 0} \lambda^N E^{[\Psi]}[\lambda^2 \bar{g},\bar{\phi},\lambda^{-1}\bar{A}],
\end{equation}
where the integer $N$ is chosen in each theory in order to ensure that the limit exists and is non-trivial.  Finite-$\lambda$ satisfaction of the field equations (for $r > \lambda \bar{R}$) then implies
\begin{equation}\label{Ebar0b}
\bar{E}^{[\Psi](0)}[\bar{g}^{(0)},\bar{\phi}^{(0)},\bar{A}^{(0)}] = 0, \qquad (\bar{r} > \bar{R}).
\end{equation}
We will refer to $\bar{E}^{[\Psi](0)}$ as the near-zone exterior field operator.  The near-zone fields obey this equation for sufficiently large $\bar{r}$, while for $\bar{r}<\bar{R}$ the fields are not constrained by the formalism (indeed, we have not even defined $\Psi^{(0)}$ in this region).  Instead, the formalism makes reference only to properties of the exterior fields $\Psi^{(0)}$ (through the charges associated with the $1/\bar{r}$ falloff at large $\bar{r}$).  In this way we obtain results independent of the detailed composition of the body (depending only on the charges).  However, in order to apply these results one must determine the exterior fields and charges associated with a given body type, which necessarily involves a continuation of $\Psi^{(0)}$ into the interior $\bar{r}<\bar{R}$.\footnote{We remind the reader that $\bar{R}$ does not represent the radius of the body, but is simply an arbitrary number designed to ensure that we impose the vacuum field equations only outside the body.}

How should this continuation be done?  The details will depend on the particular theory under consideration, but a basic guide is the physical interpretation of the large-$\bar{r}$ region as representing the ``buffer zone,'' far enough from the body that its fields may be approximated as monopole ($1/\bar{r}$) fields, but still close enough that the external universe's fields may be approximated as constants.  In the physical (finite $\lambda$) solution, equation \eqref{Ebar0b} will be approximately satisfied in this zone, and in particular the approximation will break down as one approaches the body.  To mimic this behavior in our perturbative solution, one should model the body interior and surrounding region using the full, non-electrovacuum field equations and, at some distance away, attempt to interpolate to a solution of the restricted equations \eqref{Ebar0b} that has the asymptotics of equations \eqref{gbar0}-\eqref{Abar0}.\footnote{Note that the full field equations may not even have solutions satisfying these asymptotically flat boundary conditions, indicating the general necessity of doing this kind of transition for modeling an isolated body.}  This ensures that the properties of the near-zone fields for large $\bar{r}$ accurately reflect the buffer zone properties of the body one desires to model.

The reason for the necessity of this awkward interpolation is that our mathematical description maps the physically finite buffer region to the infinite coordinate region of sufficiently large $\bar{r}$.  This allows the charges to be precisely defined in terms of asymptotic $\bar{r} \rightarrow \infty$ behavior, but at the cost of the satisfaction of the full field equations, so that the interpolation described above is required.  An alternative, physically equivalent approach would be to consider solutions to the full field equations, but evaluate the surface integral formulae only in a buffer region.  The reader that prefers this viewpoint may regard all of our surface integral formulae as holding for approximate spheres in an approximate buffer zone.

If the theory is invariant under the scalings of equation \eqref{Ebar}, then these subtleties do not arise.  To determine the near-zone fields of a body one simply considers the full non-electrovacuum field equations coupled to the matter model of choice, demanding stationarity and the asymptotic boundary conditions given by equations \eqref{gbar0}-\eqref{Abar0}.

\subsubsection{Summary of Consequences of Assumptions}

Returning to the multi-index notation, we may compactly summarize the important consequences of our assumptions as 
\begin{align}
\Psi^{(0)}_I & = \hat{\Psi}_I(t) + O(r) \label{Psi0} \\
\Psi^{(1)}_I & = \frac{a_I(t,\vec{n})}{r} + O(r^0) \label{Psi1} \\
\bar{\Psi}^{(0)}_I & = \hat{\Psi}_I(t_0) + \frac{a_I(t_0,\vec{n})}{\bar{r}} + O\left(\frac{1}{\bar{r}^2} \right), \label{Psibar0}
\end{align}
where $\Psi^{(0)}$ satisfies the (electrovacuum) field equations, $\Psi^{(1)}$ satisfies the linearized (electrovacuum) field equations (off of $\Psi^{(0)}$) for $r>0$, and $\bar{\Psi}^{(0)}$ satisfies the near-zone exterior field equation, equation \eqref{Ebar0b}, for sufficiently large $\bar{r}$.  (Each $\bar{\Psi}^{(0)}$ is defined individually in \eqref{gbar}-\eqref{cowgirl}.) We will always set the ``hatted metric'' to $\eta$ by coordinate choice, i.e., $\hat{\Psi}=\{\eta,\hat{\phi},\hat{A}^\mu\}$.  We remind the reader that indices of the perturbation \eqref{Psi1} are \textit{not} raised and lowered with the background metric $g^{(0)}$, and that the near-zone background fields are always considered with natural index position $\bar{\Psi}^{(0)}_I$, rather than conjugate position $\bar{\Psi}^{(0)I}$.

\subsection{Point Particle Description}

Our above assumptions proceed entirely from considerations regarding finite, extended bodies, as spelled out in detail in \cite{gralla-wald, gralla-motion}.  However, the resulting far-zone perturbations are singular at $r=0$ (going like $1/r$), which suggests a connection---\textit{within linearized theory}---to effective point particle sources.  (Readers familiar with the distributional equality $\nabla^2(1/r)=-4 \pi \delta(\vec{x})$ may readily attest to this fact.)  This connection was exploited in \cite{gralla-wald,gralla-motion}, leading to the result that the far-zone perturbations are sourced by effective point particle sources characterized by a finite number of charges.  We now repeat/improve the derivation of this result, while going beyond the treatment previously given in that we provide a formula for the effective charges.

The far-zone perturbations, equations \eqref{g1far}-\eqref{A1far}, satisfy the linearized field equations for $r>0$,
\begin{align}
E^{[g](1)}_{\mu \nu}[g^{(1)},\phi^{(1)},A^{(1)}] & = 0 & (r>0) \ \\
E^{[\phi](1)}[g^{(1)},\phi^{(1)},A^{(1)}] & = 0 & (r>0) \ \\
E^{[A](1)\mu}[g^{(1)},\phi^{(1)},A^{(1)}] & = 0 & (r>0),
\end{align}
which in the multi-index notation reads simply
\begin{equation}\label{icecreamtruck}
E^{(1)}_I[\Psi^{(1)}] = 0 \qquad (r>0).
\end{equation}
Since the perturbations diverge only as $1/r$, however, they are locally integrable, and hence may be straightforwardly promoted to distributions defined everywhere (including $r=0$).  

To define distributions we take our space of test functions to be $C^\infty$ compact support functions and perform integrals using the natural volume element $\epsilon=\sqrt{-g} d^4x$ associated with the background metric $g^{(0)}$.  (We write this element as $\sqrt{-g}$, even though $\sqrt{-g^{(0)}}$ might be a more accurate notation.)  We define distributions with both raised and lowered multi-index.  Raised-index distributions act on lowered-index test functions $f_I=\{f^{[g]}_{\mu \nu},f^{[\phi]},f^{[A]\mu}\}$, while lowered-index distributions act on raised-index test functions.  We raise and lower indices on distributions by raising and lowering the test function using the map $q_{IJ}$ introduced in equations \eqref{lowerPsi}-\eqref{raisePsi2}.  For example, given a lowered-index distribution $\mathcal{D}_I$, the raised-index version is defined as
\begin{equation}
\langle \mathcal{D}^I,f_I \rangle = \langle \mathcal{D}_I,q^{IJ}f_J \rangle,
\end{equation}
where the notation $\langle a,b \rangle$ indicates the action of a distribution $a$ on a test function $b$.  This convention guarantees consistency with the convention of raising and lowering ordinary functions with $q_{IJ}$, and allows us to blur the distinction between raised and lowered index distributions.  Thus we will often write $\langle \mathcal{D},f \rangle$ instead of $\langle \mathcal{D}_I,f^I \rangle$ or  $\langle \mathcal{D}^I,f_I \rangle$.

Returning to the task at hand, we associate the distribution $\tilde{\Psi}^{(1)}$ with the locally integrable function $\Psi^{(1)}$ by 
\begin{equation}
\langle \tilde{\Psi}^{(1)},f \rangle  = \int \sqrt{-g} d^4 x \ f_I \Psi^{(1)I}.
\end{equation}
Since the linearized field equations are linear, their operation on a distribution is automatically defined by
\begin{equation}
\langle E^{(1)}[\tilde{\Psi}^{(1)}], f \rangle  = \int \sqrt{-g} d^4 x \ E^{(1)}_I[f] \Psi^{(1)I},
\end{equation}
where the self-adjointness of $E^{(1)}_I$ has been used (see remarks below equation \eqref{keystone}).  Since $E_I^{(1)}$ acts on the smooth function $f$, the integral is well-defined, and we may equivalently write
\begin{equation}
\langle E^{(1)}[\tilde{\Psi}^{(1)}], f \rangle = \lim_{\epsilon \rightarrow 0} \int_{r>\epsilon} \sqrt{-g} d^4x E^{(1)}_I[f]\Psi^{(1)I} .
\end{equation}
However, since the integral now ranges only over points where $r>0$, we may make use of the fact that the perturbations $\Psi^{(1)}$ satisfy the linearized field equations on that domain.  To do so, we use equation \eqref{keystone} to ``move the field operator from the test function to the field'', giving
\begin{align}
\langle & E^{(1)}[\tilde{\Psi}^{(1)}], f \rangle = \lim_{\epsilon \rightarrow 0} \int_{r>\epsilon} \sqrt{-g} d^4x \nonumber \\ & \times \Big( E^{(1)}_I[\Psi^{(1)}]f^I + \nabla_a \omega^a[\Psi^{(0)},f,\Psi^{(1)}] \Big).
\end{align}
The first term vanishes by equation \eqref{icecreamtruck}, while the second may be converted to a surface integral by Stokes theorem,
\begin{equation}\label{okay-panda}
\langle E^{(1)}[\tilde{\Psi}^{(1)}], f \rangle = \lim_{\epsilon \rightarrow 0} \int_{r=\epsilon} \sqrt{-h} d^3x \Big( N_a \omega^a[\Psi^{(0)},f,\Psi^{(1)}] \Big),
\end{equation}
where $\sqrt{-h} d^3x$ is the 3-volume element associated with the induced metric $h$ on the surface $r=\epsilon$ (a timelike world tube for sufficiently small $\epsilon$), and $N_a$ is the normal vector.  To proceed further we adopt our locally Minkowskian coordinate system, equation \eqref{g0}, where we have $g_{\mu \nu}=\eta_{\mu \nu}+O(r)$.  In particular we have $\sqrt{-h}d^3x = r^2 (1+O(r))d\Omega dt$, $N_0=O(r)$, and $N_i=-n_i+O(r)$.  Furthermore, from the form of $\omega^a$ (equation \eqref{omegaform}) together with the fact that $\Psi^{(0)}$ and $f$ are smooth, we have
\begin{equation}\label{omegaf}
\omega^i = f^J \mathbf{F}^{ij}_{IJ}[\Psi^{(0)}] \partial_{j} \Psi^{(1)I} + O(1/r).
\end{equation}
Thus equation \eqref{okay-panda} becomes
\begin{align}
\langle & E^{(1)}[\tilde{\Psi}^{(1)}], f \rangle = - \lim_{r \rightarrow 0} \int r^2 d\Omega dt \ n_i \omega^i[\Psi^{(0)},f,\Psi^{(1)}] \label{fish}\\
& = - \int dt (f^J \mathbf{F}^{ij}_{IJ})|_{\gamma} \lim_{r \rightarrow 0} \int r^2 d\Omega \ n_i \partial_{j} \Psi^{(1)I} \\
& = - \int dt (f^J \mathbf{F}^{ij}_{IJ})|_{\gamma} \int d\Omega \ n_i \left(\frac{\partial a^I}{\partial n^j} - n_j a^I \right),
\end{align}
where in the last line we have used the form of the perturbations, equation \eqref{Psi1}.  However, since the action of $E^{(1)}[\tilde{\Psi}^{(1)}]$ depends only on the test function evaluated at $r=0$ (i.e., on $\gamma$), the distribution $E^{(1)}[\tilde{\Psi}^{(1)}]$ is in fact a delta function,
\begin{equation}\label{eff-delta}
E^{(1)}_I[\tilde{\Psi}^{(1)}] = N_I(t) \delta^{(3)}(x^i),
\end{equation}
with
\begin{align}\label{swallow}
N_J = 4 \pi \mathbf{F}^{ij}_{IJ}[\Psi^{(0)}]\big|_\gamma \Big\langle \Big\langle n_i n_j a^I - n_i \frac{\partial a^I}{\partial n^j} \Big\rangle \Big\rangle,
\end{align}
where the double angle brackets represent an average over $\vec{n}$.
Once the symplectic current (and hence $\mathbf{F}^{ij}_{IJ}[\Psi]$) is known, equation \eqref{swallow} gives an explicit formula for the coefficients of the delta function effective source in terms of the background solution $\Psi^{(0)}$ and the coefficients $a^I(t,\vec{n})$ of the $1/r$ divergence of the perturbations $\Psi^{(1)}$.  Since these coefficients may be determined in the near-zone (see equation \eqref{Psibar0}), equation \eqref{swallow} contains a prescription for determining the charges of a body from its fields.  In practice, however, it is easier to work directly with the symplectic current $\omega^a$ rather than the function $\mathbf{F}^{ij}_{IJ}$ that was defined in equation \eqref{omegaform}.  In light of equation \eqref{fish} and the antisymmetry of $\omega^a$ we may rewrite equation \eqref{swallow} as
\begin{align}\label{hummingbird}
N_J(t) = \lim_{r \rightarrow 0} \int r^2 d\Omega \ n_i \omega^i\left[ \Psi^{(0)},\frac{a^I(t,\vec{n})}{r},f^I=\delta^I_{\ J}\right],
\end{align}
where the notation $f^I=\delta^I_{\ J}$ indicates that, in determining the $J$th component of $N$, one should choose the test function $f^I$ which equals zero in all components except the $J$th, where it equals one.\footnote{Although a good test function must be of compact support, the integral in equation \eqref{hummingbird} ranges over only a finite domain and one may instead work with the function $f^I=\delta^I_{\ J}$.}  Note that the index position is important here; it is required that the raised-index components $f^I$ be assigned these values (i.e., $\delta_1 \Psi^I = \delta^I_{\ J} \neq \delta_1 \Psi_I$), corresponding to the ``conjugate'' positions of the field values, $f^I=\{f^{[g]\mu \nu},f^{[\phi]}, f^{[A]}_{\ \ \ \mu}\}$.  Equation \eqref{hummingbird} gives a convenient prescription for determining the charges from the background (far-zone) solution $\Psi^{(0)}_I$ and the coefficients $a^I$, which may be determined from the near-zone field configuration, equation \eqref{Psibar0}, corresponding to the body of interest.  Note that in many cases of interest the symplectic current will depend only on $\hat{\Psi} = \Psi^{(0)}|_\gamma$ (i.e., there is no dependence on derivatives of the background configuration), which also appears in the near-zone field configuration.  In this case equation \eqref{hummingbird} may always be rewritten as a formula making reference only to the near-zone solution $\bar{\Psi}^{(0)}$.

\subsection{Application of Bianchi Indentity and Derivation of Formulae for Motion and Charges}\label{sec:usebianchi}

Equation \eqref{eff-delta} shows that the first-order perturbations have effective delta-function source.  We may gain significantly new information about this source through the use of the Bianchi identity for the theory.  To proceed further we restrict explicitly to a theory of a single scalar and vector field (in addition to the metric), although it is straightforward to continue in total generality (see section V of paper I).  After linearization, the Bianchi identity \eqref{bianchi-identity} becomes
\begin{align}\label{bianchi-lin}
\nabla^a E^{[g](1)}_{ab} & - \frac{1}{2} \nabla_a E^{[A](1)a} A^{(0)}_b \nonumber \\ & = E^{[A](1)a} \nabla_{[a} A^{(0)}_{b]} - \frac{1}{2} E^{[\phi](1)} \nabla_b \phi^{(0)},
\end{align}
where $\nabla$ is the derivative operator associated with the background metric $g^{(0)}$.  This identity is linear in the perturbations $\Psi^{(1)}$ and therefore holds automatically on the distributional versions $\tilde{\Psi}^{(1)}$.  In particular, applying the identity to equation \eqref{eff-delta} indicates that solutions will exist only when
\begin{align}\label{bianchi-N}
\nabla^\mu \big( N^{[g]}_{\mu \nu} \delta(x^i) \big) & - \frac{1}{2} \nabla_\mu \big( N^{[A]\mu} \delta(x^i) \big) A^{(0)}_\nu \nonumber \\ & = \left( N^{[A]\mu} \nabla_{[\mu} A^{(0)}_{\nu]} - \frac{1}{2} N^{[\phi]} \nabla_\nu \phi^{(0)} \right) \delta(x^i),
\end{align}
where we have dropped the superscript $(3)$ on the delta function for readability.  Working out the consequences of this equation is simplified by use of Fermi normal coordinates (e.g., \cite{poisson-review}) about the worldline $x^i=0$.  In these coordinates the metric takes the form
\begin{equation}
g^{(0)}_{\mu \nu} = \eta_{\mu \nu} - t_\mu t_\nu a_i(t) x^i + O(r^2),
\end{equation}
where $t_\mu=(-1,0,0,0)$ and $a^i(t)$ are the spatial components of the four-acceleration of the worldline, $a^\mu=(0,a^i)$.  The Christoffel symbols in the Fermi coordinate system are given by
\begin{equation}
\Gamma^i_{00} = \Gamma^0_{i0} = a^i + O(r),
\end{equation}
with all other symbols vanishing on the worldline (i.e., $\Gamma^\mu_{\nu \rho} = O(r)$ except as indicated above).  In terms of the Fermi coordinates equation \eqref{bianchi-N} becomes
\begin{align}
0 & = \delta(\vec{x}) \Big\{ -2 \partial_0 N^{[g]}_{00} + 2 a^i N^{[g]}_{i0} - \partial_0 N^{[A]0} A_0 \nonumber \\ & \qquad \qquad - a_i N^{[A]i} A_0  + N^{[\phi]} \partial_0 \phi - 2 N^\alpha \partial_{[\alpha} A_{0]} \Big\} \nonumber \\ & \quad + \partial^i \delta(\vec{x}) \Big\{ 2N^{[g]}_{i0} - N^{[A]}_i A_0 \Big\} \label{relat1} \\
0 & = \delta(\vec{x}) \Big\{ -2 \partial_0 N^{[g]}_{0i} + 2 a^j N^{[g]}_{ij} + 2 a_i N^{[g]}_{00} - \partial_0 N^{[A]0} A_i \nonumber \\ & \qquad \qquad - a_j N^{[A]j} A_i + N^{[\phi]} \partial_i \phi - 2 N^{[A]\alpha}\partial_{[\alpha} A_{i]} \Big\} \nonumber \\ & \quad + \partial^j \delta(\vec{x}) \Big\{ 2 N^{[g]}_{ij} - N^{[A]}_{\ \ \ j} A_i \Big\}, \label{relat2}
\end{align}
where the superscript $(0)$ on the background fields $\phi^{(0)}$ and $A^{(0)}$ has been dropped for readability.  To determine the implications of these relations we integrate against a test function and use the fact that its value and derivative are arbitrary.  This shows that the terms proportional to $\partial_j \delta(\vec{x})$ must separately vanish, while the terms proportional to $\delta(\vec{x})$ must equal the spatial divergence of those proportional to  $\partial_j \delta(\vec{x})$.  From the vanishing of the $\partial_j \delta(\vec{x})$ terms we have
\begin{align}\label{llama}
2 N^{[g]}_{i\mu} = N^{[A]}_{i} A_{\mu},
\end{align}
which provides constraints on the charges $N_I$.  In particular, the anti-symmetric part of equation \eqref{llama} shows that $N^{[A]}_{\ \ \ i}$ must be proportional to $A^{(0)}_i$, and the symmetric part then determines $N^{[g]}_{i \mu}$ in terms of the constant of proportionality.  We thus have for some $\chi(t)$ that
\begin{align}
N^{[A]}_i & = \chi A_i \label{Niprop} \\
N^{[g]}_{i\mu} & = \frac{1}{2} \chi A_i A_\mu, \label{Nijprop}
\end{align}
where we remind the reader that superscript $(0)$'s have been dropped.  Thus the Bianchi identity in fact reduces the number of independent charges from fifteen ($N_I=\{N^{[g]}_{\mu \nu}, N^{[A]\mu}, N^{[\Phi]}\}$) to just four, which may be taken to be $N^{[g]}_{00}$, $N^{[A]0}$, $\chi$, and $N^{[\phi]}$.  We give these charges the suggestive names,
\begin{align}
N^{[g]}_{00} & = \frac{1}{2} M  \label{Mname} \\
N^{[\phi]} & = - q \label{qname} \\
N^{[A]0} & = e \label{ename} \\ 
\chi & = \hat{e}. \label{ehatname}
\end{align}
The remaining relations in equations \eqref{relat1} and \eqref{relat2} (those proportional to the value, rather than derivative, of the test function after integration) then give
\begin{align}
\partial_0 M &= - A_0 \partial_0 e - \frac{1}{2} \hat{e} \partial_0 ( A^i A_i ) - q \partial_0 \phi \label{mass-evolution-coord} \\
M a_i & = \partial_0 e A_i - \frac{1}{2} \hat{e} (A^j A_j) + \partial_0( \hat{e} A_i A_0 ) \nonumber \\ & \qquad + 2 e \partial_{[0}A_{i]} + q \partial_i \phi, \label{eom-coord}
\end{align}
where we again drop the superscript $(0)$ on the background vector field $A^{(0)\mu}$ and scalar $\phi^{(0)}$ for readability.  Equations \eqref{mass-evolution-coord} and \eqref{eom-coord} give the mass evolution and equation of motion for the small body in Fermi normal coordinate form.  We can also rewrite equation \eqref{eff-delta} in light of equations \eqref{Niprop}-\eqref{ehatname}, giving
\begin{align}
E^{[g](1)}_{00} & = \frac{1}{2} M \delta(\vec{x}) \\
E^{[g](1)}_{i0} & = \frac{1}{2} \hat{e} A^{(0)}_i A^{(0)}_0 \delta(\vec{x}) \\
E^{[g](1)}_{ij} & = \frac{1}{2} \hat{e} A^{(0)}_i A^{(0)}_j \delta(\vec{x}) \\
E^{[\phi](1)} & = - q \delta(\vec{x}) \label{Ephi1coord} \\
E^{[A](1)0} & = e \delta(\vec{x}) \\
E^{[A](1)i} & = \hat{e} A^{(0)i} \delta(\vec{x}) \label{EAi1coord}
\end{align}
Finally, the formulae for the charges \eqref{hummingbird} become
\begin{align}
\frac{1}{2} M & = \lim_{r \rightarrow 0} \int r^2 d\Omega \ n_i \omega^i\left[ \Psi^{(0)},\frac{a^I}{r},\{ \delta^\mu_{\ 0} \delta^\nu_{\ 0},0,0\}\right] \label{Mcoord} \\
-q & = \lim_{r \rightarrow 0} \int r^2 d\Omega \ n_i \omega^i\left[ \Psi^{(0)},\frac{a^I}{r},\{ 0,1,0\}\right] \label{qcoord} \\
e & = \lim_{r \rightarrow 0} \int r^2 d\Omega \ n_i \omega^i\left[ \Psi^{(0)},\frac{a^I}{r},\{ 0,0,\delta_{\ \mu}^0 \}\right] \label{ecoord} \\
\hat{e} A^i & = \lim_{r \rightarrow 0} \int r^2 d\Omega \ n_i \omega^i\left[ \Psi^{(0)},\frac{a^I}{r},\{ 0,0,\delta_{\ \mu}^i\}\right], \label{ehatcoord}
\end{align}
where it is understood that $\hat{e}$ vanishes if $A^i$ vanishes.  (Since only the product $\hat{e}A^i$ appears in our equations of motion, we may as well make this simplification.)  In the case that $A^i$ does not vanish, we can provide an explicit expression for $\hat{e}$ by dotting with $A_i$ and using the form \eqref{omegaf}, giving
\begin{equation}
\hat{e} = \lim_{r \rightarrow 0} \int r^2 d\Omega \ n_i \omega^i\left[ \Psi^{(0)},\frac{a^I}{r},\{ 0,0,\frac{\delta^i_{\ \mu} A_i}{A^j A_j}\}\right]. \label{ehatcoord2}
\end{equation}
We remind the reader that in equations \eqref{Mcoord}-\eqref{ehatcoord2} the index position is important  for the third argument of $\omega^a$; the symplectic current is viewed as a function of raised-multi-index variations $\delta_1 \Psi^I$ and $\delta_2 \Psi^I$, and one must send $\delta_2 \Psi^I$ (rather than $\delta_2 \Psi_I$) to the prescribed values above.  For example, to compute the mass, one sends $\delta_2 g^{\mu \nu} = \delta^\mu_{\ 0} \delta^\nu_{\ 0}$, corresponding to $\delta_2 g_{\mu \nu} = -\delta_\mu^{\ 0} \delta_\nu^{\ 0}$.

\subsection{Covariant Presentation of Results}
Equations \eqref{mass-evolution-coord}-\eqref{ehatcoord2} constitute our results for general theories, given in Fermi coordinate form.  We now rewrite these results in covariant form, in terms of the four-velocity $u^a$ of the worldline and the projection orthogonal, $P_{ab}=g_{ab}+u_a u_b$.  We have for the motion that
\begin{align}
\Big[ M - \hat{e} & \left( u^b A_b \right)^2 \Big] u^b \nabla_b u^a = P^{ab} \Big\{ u^c \nabla_c e A_b \nonumber \\ -& \frac{1}{2} \hat{e} \nabla_b \left( P_{cd} A^c A^d \right) + u^c \nabla_c \left( \hat{e} u^d A_d A_b \right) \nonumber \\ & + 2 e u^c \nabla_{[c} A_{b]} + q \nabla_b \phi \Big\}, \label{eomcov}
\end{align}
for the mass evolution that
\begin{align}
u^a \nabla_a M = u^a \left\{ -u^b \nabla_b e A_a + \frac{1}{2} \hat{e} \nabla_a \left( P_{bc}A^b A^c \right) - q \nabla_a \phi \right\},\label{massevcov}
\end{align}
 for the linearized fields that
\begin{align}
E^{[\phi](1)} & = - \int_\gamma q \ \delta_4(x,z(\tau)) d\tau \label{Ephi1cov} \\ 
E^{[A](1)a} & = \int_\gamma \left( u^a \left[ e + \hat{e} (A_c u^c) \right] + \hat{e} A^a \right) \delta_4(x,z) d\tau \label{EA1cov}  \\
E^{[g](1)}_{ab} & = \frac{1}{2} \int_\gamma \left( u_a u_b \left[ M - \hat{e} (A_c u^c)^2 \right] + \hat{e} A_a A_b \right) \nonumber \\ & \qquad \qquad \times \delta_4(x,z) d\tau, \label{Eg1cov}
\end{align}
and for the charges that
\begin{align}
M & = 2 \lim_{r \rightarrow 0} \int r^2 d\Omega \ n_i \omega^i\left[ \Psi^{(0)},\frac{a^I}{r},\{ u^\mu u^\nu,0,0\}\right] \label{Mcov} \\ 
q & = - \lim_{r \rightarrow 0} \int r^2 d\Omega \ n_i \omega^i\left[ \Psi^{(0)},\frac{a^I}{r},\{ 0,1,0 \}\right] \label{qcov} \\
e & = \lim_{r \rightarrow 0} \int r^2 d\Omega \ n_i \omega^i\left[ \Psi^{(0)},\frac{a^I}{r},\{ 0,0,-u_\mu\}\right] \label{ecov} \\
\hat{e} & = \lim_{r \rightarrow 0} \int r^2 d\Omega \ n_i \omega^i\left[ \Psi^{(0)},\frac{a^I}{r}, \{ 0,0,\tilde{A}^\perp_\mu\}\right], \label{ehatcov}
\end{align}
where $\tilde{A}^\perp_a=P_{ab}A^b/(P^{cd}A_c A_d)$ is a vector pointing along the spatial projection of $A^a$.  (It is understood that if spatial projection vanishes, $P_{ab}A^b=0$, then the charge $\hat{e}$ vanishes and equation \eqref{ehatcov} does not apply.)  In all of these equations we have dropped superscript $(0)$ on background fields.  While equations \eqref{eomcov}-\eqref{Eg1cov} are fully covariant, we have left equations \eqref{Mcov}-\eqref{qcov} in a form valid only in local inertial coordinates, $g^{(0)}_{\mu \nu}=\eta_{\mu \nu}+O(r)$.  (Note in particular that $u^\mu =(1,0,0,0)$ and $n^i=x^i/r$.)  Since any computation of the charges will likely use such coordinates, we see little reason to covariantize further.  We remind the reader that index position is important here; see discussion below equation \eqref{ehatcoord2}.  For example, in computing the mass one must send $\delta_2 g^{\mu \nu} = u^\mu u^\mu$, corresponding to $\delta_2 g_{\mu \nu} = - u_\mu u_\mu$.  

We now indicate how the results \eqref{eomcov}-\eqref{ehatcov} generalize to theories containing additional fields besides the single scalar and vector field considered here.  First, it was shown in paper I that analogs of equations \eqref{eomcov}-\eqref{Eg1cov} exist for a theory with arbitrary types of tensor fields, though the explicit form was not obtained.  As for the formulae for the charges, new to this paper, our general formula \eqref{hummingbird} remains valid for a theory with arbitrary types of tensor fields, where the multi-index runs over each component of each field.  However, the application of the Bianchi identity to simplify this formula, leading to a smaller number of charges (just four in the scalar-vector-tensor case), must be redone to determine analogs of \eqref{Mcov}-\eqref{ehatcov} for a general theory.  Thus, for an arbitrary theory, the status is that analogs of our results \eqref{eomcov}-\eqref{ehatcov} are known to exist, but would take a complicated form that has not yet been written down.

An exception to this statement, however, is the case where the theory conditions additional scalar fields only.  Following through the calculations of section \ref{sec:motion} for a theory of a metric, a vector field, and a collection of scalar fields $\phi_A$, it is easy to establish that equations \eqref{eomcov}-\eqref{ehatcov} generalize as follows.  First, there is a new scalar charge $q_A$ associated with each field $\phi_A$.  To generalize equations \eqref{eomcov} and \eqref{massevcov}, one simply adds an index $A$ to each incidence of the scalar field $\phi$ and its charge $q$, turning these terms into sums over the collection of fields.  For example, the last term of equation \eqref{eomcov}, $q P^{ab} \nabla_b \phi$, becomes $q_A P^{ab} \nabla_b \phi_A$, where there is a sum over $A$.  Equation \eqref{Ephi1cov} is modified by creating an additional copy for each field: one adds an index $A$ to the field $\phi$, charge $q$, and field equation operator $E^{[\phi](1)}$.  Equations \eqref{EA1cov} and \eqref{Eg1cov} are unmodified.  Finally, there is a copy of equation \eqref{qcov} for each new charge, where the second argument of the symplectic current contains a non-zero entry only for the slot associated with the field whose charge is under consideration.  Equations \eqref{Mcov}, \eqref{ecov}, and \eqref{ehatcov} are unmodified, except in the trivial sense that the symplectic current has additional arguments for the additional scalar fields, into which zeros are to be inserted.

\subsection{Usage of Results}

We now summarize how the results of this section may be applied in a specific theory.  First one must obtain the near-zone fields for the body type of interest as a function of internal parameters (e.g., baryon number, central pressure, etc.) and the local external field values, as described in section \ref{sec:nonscaleinv}.  One may use any asymptotically Minkowskian coordinate system $(t,x^i)$.  Next extract the asymptotic $1/r$ fields in that coordinate system, i.e., determine the $a/r$ terms defined by equations \eqref{gbar0}-\eqref{Abar0}.  Then plug the asymptotic fields into equations \eqref{Mcov}-\eqref{ehatcov} with $u^\mu=(1,0,0,0)$ to determine the charges of the body as a function of the internal parameters and external field values.  (See also equations \eqref{Mcoord}-\eqref{ehatcoord}, where coordinate components are written out explicitly.)  After selecting a background spacetime, plug the charge functions into equations \eqref{eomcov} and \eqref{massevcov}, where the value of the background configuration $\{g,A,\phi\}$ on the worldline is used as the external field value for the charges.  For bodies with a single internal parameter, equation \eqref{massevcov} will fix the evolution of this parameter, and the coupled system may be solved for the motion.  For bodies with more internal parameters, evolution laws for the parameters must be determined by physical arguments or by finding near-zone solutions with slowly varying asymptotic fields.  However, once evolution laws (consistent with \eqref{massevcov}) for the paramters are fixed, one may solve equation \eqref{eomcov} to determine the motion.  The motion and charge evolution may then be plugged in to equations \eqref{Ephi1cov}-\eqref{Eg1cov} to determine the perturbation caused by the body.  Note that it may be convenient to eliminate the step of explicitly extracting the $a/r$ near-zone fields by rewriting the surface integrals in the near-zone, as done for each specific theory treated below.

\section{General Relativity with a Cosmological Constant}\label{sec:GRCC}

We now work out the consequences of the general prescription for mass, charge and motion in the example of general relativity with a cosmological constant.  The Lagrangian is  
\begin{equation}\label{LGR}
L = \frac{1}{16\pi}\left( R - 2 \Lambda \right),
\end{equation}
which gives field equation operator
\begin{equation}\label{EGR}
16 \pi E^{[g]}_{ab}[g] = G_{ab} + \Lambda g_{ab}
\end{equation}
and symplectic potential
\begin{equation}\label{ThetaGR}
16 \pi \Theta^a[g,\delta g] = 2 g_{bc} \nabla^{[a}\delta g^{b]c} .
\end{equation}
The symplectic current is then given by
\begin{align}\label{omegaGR}
32 \pi \omega^a[g,h,j] & = j^{ab}\nabla_b h + j^{bc} \nabla^a h_{bc} - 2 j^{bc} \nabla_b h_{c}^{\ a} \nonumber \\ & - j \nabla^a h + j \nabla_b h^{ab} - h \leftrightarrow j.
\end{align}
Here we have set $h^{ab} = \delta_1 g^{ab}$ and $j^{ab} = \delta^2 g^{ab}$, and we raise and lower indices on these quantities with the background index, so that $h_{ab} = - \delta_1 g_{ab}$ and $j_{ab}=-\delta_2 g_{ab}$. Since there are no fields besides the metric, the only nonvanishing charge is the mass $M$.  To determine a formula for this charge, equation \eqref{Mcov} (equivalently \eqref{Mcoord}) instructs us to consider the symplectic current with $h^{\mu \nu} \rightarrow a^{\mu \nu}/r$ and $j^{\mu \nu} \rightarrow u^\mu u^\nu = \delta^\mu_{\ 0} \delta^\nu_{\ 0}$.  (We have dropped the superscript $[g]$ on $a^{[g]}_{\mu \nu}$ since our theory has no additional fields.  Note also that $h_{\mu \nu}=-a_{\mu \nu}$, since we do not use the background metric to raise and lower indices of $a_{\mu \nu}$.)  Computing $\omega^i$ and keeping only the most singular ($1/r^2$) terms gives
\begin{align}\label{omegaMsingGR}
32 \pi \omega_i\left[g,\frac{a^{\mu \nu}}{r},u^\mu u^\nu\right] & = \partial_k \left( \frac{a_{ki}}{r} \right) - \partial_i \left( \frac{a_{kk}}{r} \right) + O\left(\frac{1}{r}\right).
\end{align}
The mass is then given by equation \eqref{Mcov} to be 
\begin{equation}
M = \frac{1}{16\pi} \lim_{r \rightarrow 0} \int r^2 d\Omega \ n^i \left\{ \partial_k \left( \frac{a_{ki}}{r} \right) - \partial_i \left( \frac{a_{kk}}{r} \right) \right\}.
\end{equation}
Taking into account the appearance of $a_{\mu \nu}$ in the near-zone background (equation \eqref{gbar0}), we may equivalently write
\begin{equation}\label{MGR}
M = \frac{1}{16\pi} \lim_{\bar{r} \rightarrow \infty} \int \bar{r}^2 d\Omega \ n^i \left( \partial_k \bar{g}^{(0)}_{ki} - \partial_i \bar{g}^{(0)}_{kk} \right),
\end{equation}
which is recognized as the famous ADM mass formula \cite{arnowitt-deser-misner}.  Thus, the charge $M$ is nothing but the ADM mass of the near-zone background spacetime.  It is notable that the ADM formula for $M$ emerged, rather than some other formula (such as that of Komar \cite{komar}) that is equivalent given the stationary field equations.  Since the ADM approach uses the Hamiltonian formulation, which itself is intimately connected to symplectic structure (e.g., \cite{iyer-wald}), one might suspect a deep and general connection between that approach and ours.  However, we have found no such connection, and it may be pure coincidence that the ADM formula emerges.

To determine the mass $M$ of a given body using equation \eqref{MGR}, one must determine the near-zone background metric.  This procedure was discussed at the conclusion of section \ref{sec:framework}.  The first step is to identify the near-zone exterior field operator.  Using the scale-invariance of the Einstein tensor, at finite $\lambda$ we have
\begin{equation}
0=G_{ab}[g] + \Lambda g_{ab} = G_{ab}[\bar{g}] + \lambda^2 \Lambda \bar{g}_{ab}.
\end{equation}
Letting $\lambda \rightarrow 0$ in scaled coordinates, we then obtain
\begin{equation}\label{extGR}
G_{\bar{\mu} \bar{\nu}}[\bar{g}^{(0)}] = 0 \qquad (\textrm{large } \bar{r}),
\end{equation}
which is satisfied for $\bar{r}$ larger than some $\bar{R}$ (see equation \eqref{Ebar0b} and discussion below).  Thus the near-zone background metric is a solution of the \textit{vacuum} Einstein equations in the large-$\bar{r}$ region.  Equation \eqref{gbar0} indicates that we should seek stationary, asymptotically flat solutions; and, following the discussion of section \ref{sec:nonscaleinv}, these solutions should be matched to interior solutions relevant for the particular matter model under consideration.  The interior solutions should satisfy the full non-electrovacuum field equations,
\begin{equation}\label{intGR}
G_{\bar{\mu} \bar{\nu}}[\bar{g}^{(0)}] + \Lambda \bar{g}^{(0)}_{\bar{\mu} \bar{\nu}} = 8 \pi T_{\bar{\mu}\bar{\nu}}, \qquad (\textrm{interior})
\end{equation}
where $T_{ab}$ is the stress-energy tensor of the matter fields.  

Thus, after choosing a matter model one should attempt to find a solution of equation \eqref{intGR} that interpolates in a reasonably smooth way to a stationary, asymptotically flat solution of \eqref{extGR}.  This provides the near-zone background metric for the body under consideration, whose mass $M$ may then be computed by equation \eqref{MGR}.  (This is just the ADM mass of the solution.)  Note, however, that while in principle the cosmological constant term in equation \eqref{intGR} should be included, in practice its contribution should be negligibly small.  This is because the requirement of a region surrounding the body where the solutions match approximately implies that the body must be small compared to the scale set by the cosmological constant, in which case its effects should be negligible everywhere in the near-zone.  Thus one may simply consider the field equations of ordinary general relativity (with no cosmological constant) when determining the mass of a particular body type.  In more general theories, however, this type of argument may not be possible, and one must in principle use the full field equations near the body.

Having interpreted the charge $M$, we may now make use of the results for the motion and metric perturbation, equations \eqref{eomcov}-\eqref{qcov}.  In particular, the motion is geodesic,
\begin{equation}\label{eomGR}
u^b \nabla_b u^a = 0,
\end{equation}
the mass is conserved,
\begin{equation}\label{massevGR}
u^a \nabla_a M = 0,
\end{equation}
and the linearized perturbation $h$ is sourced by a point particle of mass $M$,
\begin{equation}\label{E1GR}
G_{ab}^{(1)}[h] + \Lambda h_{ab} = \frac{M}{8\pi}\int_\gamma u_a u_b \delta_4(x,z) d\tau.
\end{equation}

To summarize, the prescription for general relativity with a cosmological constant is as follows.  First, specify what type of body is being considered by giving a matter model and determining the relevant stationary, asymptotically flat solution to equations \eqref{intGR} and \eqref{extGR} (the cosmological constant term in \eqref{intGR} may be neglected).  Compute the ADM mass of this spacetime and assign this value to the parameter $M$.  Given a background solution of the vacuum Einstein equation with cosmological constant, the motion of that body will be geodesic, and the perturbations produced will satisfy equation \eqref{E1GR}.

\section{Scalar-Tensor Theories}\label{sec:scalar-tensor}

For a theory of the metric and a scalar field, \eqref{massevcov}-\eqref{eomcov} become
\begin{align}
u^a \nabla_a M & = -q u^a \nabla_a \phi^{(0)} \label{massevES} \\
M u^b \nabla_b u^a & = q \left( g^{(0)ab} + u^a u^b \right) \nabla_b \phi^{(0)}. \label{eomES}
\end{align}
The main new subtlety in the scalar-tensor case (compared to general relativity) is that the charges $q$ and $M$ are not conserved.  Indeed, the mass varies according to equation \eqref{massevES}, while the evolution of $q$ is simply not constrained.  The lack of deterministic evolution is a result of imposing field equations only outside of a worldtube: While in general relativity satisfaction of the vacuum field equations in this region suffices to determine the motion of a body, in a scalar-tensor theory we only learn that the motion is given in terms of two charges, one of which is not constrained at all by the theory.  (See paper I for more discussion of this point.)  Of course, if the region containing the body is ``reinserted'' at finite $\lambda$ for a given scalar-tensor theory and a given matter model, then one should have deterministic evolution at finite $\lambda$ and hence deterministic evolution within our perturbation series.  The precise deterministic evolution will depend in detail on the precise matter model considered, but our results indicate that this freedom will show up only in terms of a single charge $q$ (the mass $M$ being given by equation \eqref{massevES} once known initially).  Thus, to complete the prescription of the motion of a body in a scalar-tensor theory, one must determine the value \textit{and evolution} of $q$ implied by specific matter models for specific types of bodies.

We now discuss this process.  Each scalar-tensor theory will have a different set of interior and exterior field equations, analogous to equations \eqref{intGR} and \eqref{extGR}, that the near-zone field configuration $\{\bar{g}^{(0)},\bar{\phi}^{(0)}\}$ should satisfy, with a smooth interpolation in an intermediate region.  Comparing equations \eqref{g0}-\eqref{phi0} and \eqref{gbar0}-\eqref{phibar0}, we see that the $\bar{r} \rightarrow \infty$ boundary conditions are that the metric must approach the Minkowski metric and the scalar field must approach the value of the far-zone scalar field evaluated at the current position of the particle.  That is, for a given time $t_0$ we seek stationary ($\bar{t}$-independent) solutions of the near-zone equations such that
\begin{align}
\bar{g}^{(0)}_{\bar{\mu}\bar{\nu}}(t_0;\bar{x}^i) & \rightarrow \eta_{\mu \nu} & (\bar{r} \rightarrow \infty) & \label{g0bndryES} \\
\bar{\phi}^{(0)}(t_0;\bar{x}^i) & \rightarrow \hat{\phi}(t_0) \equiv \phi^{(0)}|_{t=t_0,x^i=0} \ \  & (\bar{r} \rightarrow \infty) & . \label{phi0bndryES}
\end{align}
How does one determine the appropriate such solution?  The simplest method is to specify a matter model together with physical assumptions (such as a fixed total baryon number) such that the solution will be uniquely determined for each possible external field $\hat{\phi}$.  In this case the field configuration $\{\bar{g}^{(0)},\bar{\phi}^{(0)}\}$ becomes a deterministic function of the local external field value $\hat{\phi}$.  The mass and charge are then also determined uniquely as functions of $\hat{\phi}$ by equations \eqref{MES} and \eqref{qES}, and we write
\begin{align}
M = M(\hat{\phi}) \label{dog}\\
q = q(\hat{\phi}) \label{cat}.
\end{align}
In this case we have $u^a \nabla_a M = M'(\phi^{(0)}) u^a \nabla_a \phi^{(0)}$ and equation \eqref{massevES} implies that
\begin{equation}\label{MqES}
M'(\hat{\phi}) = - q(\hat{\phi}).
\end{equation}
Equation \eqref{MqES} gives a constraint that must be satisfied by any valid prescription for obtaining the charges $\{M,q\}$ as a function of the external field value $\hat{\phi}$.  In light of the constraint we may interpret the charge $q$ as representing the sensitivity of the mass of the body to adiabatic changes in the local external $\phi$-field value.  The notion of a sensitivity was first introduced by Eardley \cite{eardley} in the context of a phenomenological approach to motion in Jordan frame Brans-Dicke theory, based on a point particle action.  Will and coauthors have subsequently applied and expanded the approach, as summarized in \cite{will-book}.  Damour and Esposito-Farese \cite{damour-espositofarese} adapted the Eardley method to a more general class of scalar-tensor theories, and furthermore related $q$ and $M$ to the asymptotic $1/r$ behavior of the fields.  Our first-principles treatment has confirmed the appearance of a sensitivity when one considers bodies whose exterior field configuration is a strict function of the local external field value (a basic assumption of previous approaches, relaxed by ours).  The sensitivity is normally defined by $s=-d \log M / d\log \phi$, and so the relationship to our $q$ is given by
\begin{equation}\label{sq}
s = -\frac{\hat{\phi}}{M} q.
\end{equation}

We emphasize, however, that the relations \eqref{dog}-\eqref{cat} (and hence \eqref{MqES}) are not fundamental to scalar-tensor theory but rather are consequences of considering a body whose exterior fields are ``state functions'' of the of the local external field value $\hat{\phi}$.  While this is surely a reasonable assumption for many bodies of interest, there is no principle that forbids more general bodies, whose external fields may be ``path functions'' of $\hat{\phi}$ or may change independently of $\hat{\phi}$ due to internal processes.  An example of the latter would be a star that cools or otherwise changes its internal structure on a timescale relevant for motion.  (Since the stellar structure and evolution equations may be drastically modified in alternative theories, this possibility is not obviously excluded for our Universe, as described by alternative theories.)  In this case one would instead have a prescription for determining the exterior fields $\{\bar{g}^{(0)},\bar{\phi}^{(0)}\}$ as functions of both $\hat{\phi}$ and some prescribed internal dependence related to (say) a slow change in equation of state.  That is, these fields would become explicit functions of the time $t_0$ along the worldline, in addition to any dependence on external fields.  In this way the mass and charge would become functions of both $t_0$ and $\hat{\phi}$, and one must respect the more general constraint \eqref{massevES}, which becomes
\begin{equation}\label{Mq2ES}
 \frac{\partial}{\partial t_0} M(t_0,\hat{\phi}) = - q (t_0,\hat{\phi}) u^a \nabla_a \phi^{(0)}.
\end{equation}

For completeness we reproduce equations \eqref{Ephi1cov}-\eqref{Eg1cov} for the body fields in the scalar-tensor case,
\begin{align}
E^{[\phi](1)} & = - \int_\gamma q \ \delta_4(x,z(\tau)) d\tau \label{Ephi1ST} \\ 
E^{[g](1)}_{ab} & = \frac{1}{2} \int_\gamma M u_a u_b \delta_4(x,z) d\tau. \label{Eg1ST}
\end{align}

\subsection{Einstein Frame Scalar-Tensor Theories}\label{sec:ES}

We now consider general relativity with the addition of a Klein-Gordon field with potential $V(\phi)$,
\begin{equation}\label{LES}
L = \frac{R}{16 \pi} - \frac{1}{2} g^{ab} \nabla_a \phi \nabla_b \phi - V(\phi).
\end{equation}
It is well known that a large class of Lagrangians can be written in this form by a field redefinition, a procedure normally referred to as ``going to the Einstein frame''.  In doing this procedure one normally assumes that the matter was coupled to the original field variables in a standard way (``minimally coupled''), so that the coupling to the new field variables becomes non-standard.  This assumption amounts simply to a choice of non-standard matter model for the Lagrangian \eqref{LES}, so that many scalar-tensor theories are subsumed under the single Lagrangian.  In this section we derive the surface-integral formulae for the charges for the Lagrangian \eqref{LES} and  evaluate them for specific types of matter coupling.

The field equation operators for the Lagrangian \eqref{LES} are
\begin{align}
E^{[g]} & = \frac{G_{ab}}{16\pi} - \frac{1}{2} T^{[\phi]}_{ab} \label{EgES} \\
E^{[\phi]} & = g^{ab} \nabla_a \nabla_b \phi - V'(\phi), \label{EphiES}
\end{align}
where
\begin{equation}
T^{[\phi]}_{ab} \equiv  \nabla_a \phi \nabla_b \phi - \frac{1}{2} g_{ab} g^{cd} \nabla_c \phi \nabla_d \phi + V(\phi) g_{ab}
\end{equation}
is normally referred to as the stress-energy of the scalar field.  The symplectic potential is
\begin{equation}
\Theta^a = \Theta_{GR}^a - \nabla^a \phi \ \delta \phi,
\end{equation}
where $\Theta_{GR}$ is the symplectic potential for general relativity, given by equation \eqref{ThetaGR}.  The symplectic current is then given by
\begin{align}
& \omega^a[  \left\{g,\phi\right\},\left\{h,\alpha\right\},\left\{j,\beta\right\}] - \omega_{GR}^a[g,h,j] = \nonumber \\ & \ \ - \beta \left( \nabla^a \alpha + \frac{1}{2} h \nabla^a \phi + h^{ab} \nabla_b \phi \right) - \binom{h \leftrightarrow j}{\alpha \leftrightarrow \beta},
\end{align}
where we view $\omega$ as a function of $h^{ab} = \delta_1 g^{ab}$, $j^{ab}=\delta_2 g^{ab}$, $\alpha=\delta_1 \phi$ and $\beta = \delta_2 \phi$, and use the background metric $g_{ab}$ to raise and lower indices on these quantities.  (The last term on the RHS stands for an additional copy of all previous terms on the RHS, except modified by exchanging $h$ with $j$ and $\alpha$ with $\beta$.)  The formulae for the charge $M$, equation \eqref{Mcov}, instructs us let $h^{\mu \nu} \rightarrow a^{[g]\mu \nu}/r$, $j^{\mu \nu} \rightarrow u^\mu u^\nu = \delta^\mu_{\ 0} \delta^\nu_{\ 0}$, $\alpha \rightarrow a^{[\phi]}/r$ and $\beta \rightarrow 0$.  Keeping only the most singular ($1/r^2$) terms, we have
\begin{equation}
\omega^i[\{g,\phi\},\{\frac{a^{[g]}_{\mu \nu}}{r},\frac{a^{[\phi]}}{r}\},\{u_\mu u_\nu,0\} ] = \omega^i_{GR} + O\left( \frac{1}{r} \right),
\end{equation}
where $\omega^i_{GR}$ is evaluated with the same arguments as $\omega^i$ (given explicitly by equation \eqref{omegaMsingGR}).  Thus the formula for the mass $M$ is identical to that of general relativity (i.e., to the ADM mass of the near-zone spacetime).  For the scalar charge $q$, equation \eqref{qcov} instructs us to let $h^{\mu \nu} \rightarrow a^{[g]\mu \nu}/r$, $j^{\mu \nu} \rightarrow 0$, $\alpha \rightarrow  a^{[\phi]}/r$ and $\beta \rightarrow 1$.  Keeping only the most singular ($1/r^2$) terms, we have
\begin{equation}
\omega_i[\{g,\phi\},\{\frac{a^{[g]}_{\mu \nu}}{r},\frac{a^{[\phi]}}{r}\},\{0,1\} ] = \partial_i \left( \frac{a^{[\phi]}}{r} \right) + O\left( \frac{1}{r} \right).
\end{equation}
Thus the charge $q$ is given by
\begin{equation}
q = - \lim_{r \rightarrow 0} \int r^2 d\Omega n^i \partial_i \left( \frac{a^{[\phi]}}{r} \right).
\end{equation}
Taking into account the appearance of $a^{[\phi]}$ in the near-zone background field, equation \eqref{phibar0}, we may rewrite this formula in terms of the near-zone background field.  We display the result along with the (ADM) formula for $M$,
\begin{align}
M & = \frac{1}{16\pi} \lim_{\bar{r} \rightarrow \infty} \int \bar{r}^2 d\Omega \ n^i \left( \partial_k \bar{g}^{(0)}_{ki} - \partial_i \bar{g}^{(0)}_{kk} \right)  \label{MES} \\
 q & = - \lim_{\bar{r} \rightarrow \infty} \int \bar{r}^2 d\Omega \ n^i \partial_i \bar{\phi}^{(0)}. \label{qES} 
\end{align}

To determine $M$ and $q$ for a particular body we must determine the near-zone fields $\bar{g}^{(0)}$ and $\bar{\phi}^{(0)}$.  Using the scaling properties of the field operators, it is easy to see that we have
\begin{align}
G_{\bar{\mu}\bar{\nu}}[\bar{g}^{(0)}] & = 8 \pi \Big( \nabla_{\bar{\mu}} \bar{\phi}^{(0)} \nabla_{\bar{\nu}} \bar{\phi}^{(0)} \nonumber \\ & - \frac{1}{2} \bar{g}^{(0)}_{\bar{\mu}\bar{\nu}} \bar{g}^{(0)\bar{\alpha}\bar{\beta}} \nabla_{\bar{\alpha}} \bar{\phi}^{(0)} \nabla_{\bar{\beta}} \bar{\phi}^{(0)} \Big)  \label{extESg} \\
\bar{g}^{(0)\bar{\alpha}\bar{\beta}} \nabla_{\bar{\alpha}} \nabla_{\bar{\beta}} \bar{\phi}^{(0)} & = 0, \qquad \textrm{(large }\bar{r}) \label{extESphi}
\end{align}
where $\nabla_a$ is the derivative operator associated with $\bar{g}^{(0)}$.  These equations agree with the full field equations with the potential $V(\phi)$ set to zero.  On the other hand, the interior satisfies the full non-electrovacuum field equations (including the potential $V(\phi)$),
\begin{align}
G_{\bar{\mu}\bar{\nu}}[\bar{g}^{(0)}] - 8 \pi T^{[\phi]}_{\bar{\mu}\bar{\nu}}[\bar{\phi}^{(0)},\bar{g}^{(0)}] & = 8\pi  T_{\bar{\mu} \bar{\nu}} \label{intESg} \\
\bar{g}^{(0)\bar{\alpha}\bar{\beta}} \nabla_{\bar{\alpha}} \nabla_{\bar{\beta}} \bar{\phi}^{(0)} - V'(\bar{\phi}^{(0)}) & = - \sigma, \ \textrm{(interior)} \label{intESphi}
\end{align}
where $T_{ab}$ and $\sigma$ are the stress-energy and scalar charge density of the matter under consideration.  To determine the charges $q$ and $M$ one should find solutions of equations \eqref{extESg}-\eqref{intESphi} in the manner described in section \ref{sec:nonscaleinv}, and plug these solutions into equations \eqref{MES} and \eqref{qES}.  The motion is then given by equation \eqref{eomES}, and the far-zone body fields are given by \eqref{Ephi1ST}-\eqref{Eg1ST}.

\subsubsection{Ordinary Coupling: Einstein-Scalar Theory}

Consider the Lagrangian \eqref{LES} with no potential, $V(\phi)=0$, and in the case that ``ordinary matter'' is endowed with an intrinsic scalar charge density $\sigma$ in addition to its stress-energy $T_{ab}$.  (In a Lagrangian viewpoint, these would be defined by equations \eqref{Tfunc} and \eqref{sigmafunc}.)  We refer to this theory as Einstein-Scalar Theory.  We now investigate the meaning of the charges $q$ and $M$ in this theory.  We therefore seek solutions to the near-zone equations \eqref{extESg}-\eqref{intESphi}, which, given $V(\phi)=0$, reduce to equations \eqref{intESg}-\eqref{intESphi} where the sources $T_{ab}$ and $\sigma$ have compact support.  These solutions should be stationary and approach $\{\eta,\hat{\phi}\}$ at large distances (see equations \eqref{g0bndryES}-\eqref{phi0bndryES}).  

We first consider a ``Newtonian body'' in the sense that 1) only linear deviations from $\{\eta,\hat{\phi}\}$ are allowed and 2) $T_{0i}$ and $T_{ij}$ are negligible.  Following the standard approach to the Newtonian limit in general relativity, we find that a gauge may be chosen where equations \eqref{intESg}-\eqref{intESphi} become
\begin{align}
\nabla^2 \Phi & = 4 \pi \rho \label{tribble1} \\
\nabla^2 \alpha & = - \sigma, \label{tribble2}
\end{align}
with $\rho=T_{00}$ and $\nabla^2=\delta^{ij} \partial_i \partial_j$, and with the fields given in terms of the potentials $\Phi$ and $\alpha$ by
\begin{align}
g_{\mu \nu} & = \eta_{\mu \nu} - 2 \Phi \delta_{\mu \nu} \label{kirk1} \\
\phi & = \hat{\phi} + \alpha. \label{kirk2}
\end{align}
From the multipole expansion we know that the solutions to equations \eqref{tribble1}-\eqref{tribble2} that are vanishing at infinity are given by
\begin{align}
\Phi & = \left( - \int \rho d^3x \right) \frac{1}{r} + O\left(\frac{1}{r^2}\right) \label{spock1} \\
\alpha & = \left( \frac{1}{4\pi} \int \sigma d^3x \right) \frac{1}{r} + O\left(\frac{1}{r^2}\right) . \label{spock2}
\end{align}
Plugging equations \eqref{kirk1}-\eqref{spock2} into equations \eqref{MES} and \eqref{qES} then gives the charges for a Newtonian body to be
\begin{align}
M & = \int \rho d^3x \label{MESNewt} \\
q & = \int \sigma d^3x \label{qESNewt}.
\end{align}
We thus see that for weak-field bodies the mass and charge are given by volume integrals of the mass density and scalar charge density, respectively.  Note that these densities are not conserved, but are related to each other by equation \eqref{bianchi-identity-nonev}.  This equation does not fix their evolution in time, which will instead depend in detail on the matter model adopted.  Thus the Einstein-scalar theory displays a wide variety of motions of bodies, even in the weak-field limit.  Indeed, all we can conclude is that the general equation of motion, equations \eqref{massevES}-\eqref{eomES}, is obeyed for some (in general non-constant) charge $q$.

It is also straightforward to consider black holes.  The well-known ``no scalar hair'' theorem \cite{beckenstein-nohair1,beckenstein-nohair2,chase,hawking-scalar} implies that the scalar field must be constant, $\phi=\hat{\phi}$, and hence the metric must be that of the Kerr family.  In this case the mass $M$ agrees with that of the Kerr solution and the charge $q$ vanishes.  Thus the mass is conserved and the black hole moves on a geodesic.

\subsubsection{Conformal Coupling with no Intrinsic Scalar Charge: Generalized Brans-Dicke Theory}

In Brans-Dicke theory and its cousins, one assumes that matter is minimally coupled to a (``Jordan frame'') metric that is conformally related to the (``Einstein frame'') metric of our Lagrangian \eqref{LES}.  This is represented by considering a matter action of the form
\begin{equation}
S_M = \int d^4x \ \Omega^{-4}(\phi) \sqrt{-g} L_M[\psi_M,\Omega^{2}(\phi) g^{ab}],
\end{equation}
where $\psi_M$ is some matter field.  Notice that the matter action depends on $\phi$ only through the conformal metric $\Omega^{-2} g_{ab}$, which is identified with ordinary matter by the relationship
\begin{align}
T^{\textrm{ord}}_{ab} & = \frac{-2}{\sqrt{-\det ( \Omega^{-2}(\phi)g_{\mu \nu})}} \frac{\delta {S_M}}{\delta (\Omega^{2}(\phi) g^{ab})} \\ 
& = \frac{-2 \Omega^4(\phi)}{\sqrt{-g}} \frac{\delta {S_M}}{\delta (\Omega^{2}(\phi) g^{ab})},
\end{align}
where $T^{\textrm{ord}}_{ab}$ represents the stress-energy tensor of ``ordinary matter'', i.e., the stress tensor one would normally assign to a matter model within general relativity.  (This is the ``Jordan frame stress tensor''.)  The stress tensor and scalar charge density of our Einstein frame theory are then related by
\begin{align}
T_{ab} & = \frac{-2}{\sqrt{-g}} \frac{\delta S_M}{\delta g^{ab}} = \Omega^2 T^{\textrm{ord}}_{ab}\\
-\sigma & = \frac{-1}{\sqrt{-g}} \frac{\delta S_M}{\delta \phi} = \Omega^{-3} \Omega' g^{ab} T^{\textrm{ord}}_{ab},
\end{align}
and so the non-electrovacuum equations are simply
\begin{align}
G_{ab} - 8 \pi T_{ab}^{[\phi]} & = 8 \pi \Omega^2 T^{\textrm{ord}}_{ab} \\
g^{ab} \nabla_a \nabla_b \phi & = \Omega^{-3} \Omega' g^{ab} T^{\textrm{ord}}_{ab}.
\end{align}
For a weak-field body (see equations \eqref{tribble1}-\eqref{tribble2} and text above), we then have
\begin{align}
\nabla^2 \Phi & = 4 \pi \Omega^{-4} \rho \\
\nabla^2 \alpha & = - \Omega^{-5} \Omega ' \rho,
\end{align}
where $\Omega$ and its $\phi$-derivative $\Omega '$ are evaluated at the background value $\phi=\hat{\phi}$.  Here $\rho = \Omega^4 T_{00}$ is the ordinary mass density, i.e., the time-time component of $T^{\textrm{ord}}$ coordinates $\Omega x^\mu$ where the metric $\Omega^{-2} g_{ab}$ equals the Minkowski metric.  Following the procedure of the previous subsection, the charges are given by
\begin{align}
M & = \Omega^{-4} \int \rho \ d^3x = \Omega^{-1} M_B \label{MBDENewt} \\
q & = \Omega^{-5} \Omega' \int \rho \ d^3 x = \Omega^{-2} \Omega' M_B, \label{qBDENewt}
\end{align}
where $M_B = \int \rho \ d^3 (\Omega x)$ is the total (``Jordan frame'') baryonic mass, i.e., the spatial integral of the time-time component of $T^{\textrm{ord}}$ in coordinates where the where the metric $\Omega^{-2} g_{ab}$ equals the Minkowski metric.  From equations \eqref{MBDENewt} and \eqref{qBDENewt} we see that the charge-to-mass ratio of Newtonian bodies depends only on the asymptotic scalar field value,
\begin{equation}
\frac{q}{M} = \Omega^{-1} \Omega' = \frac{d}{d \phi} \log \Omega ,
\end{equation}
so that the motion of a Newtonian body is independent of its composition.  (Note that this charge-to-mass ratio is precisely that required for geodesic motion in the metric $\Omega^{-2} g_{ab}$, consistent with the fact that matter is by assumption minimally coupled to that metric.)  The constraint $d M / d\phi = -q$ shows that the baryonic mass $M_B$ is independent of the asymptotic scalar field value, in agreement with what is found in the Jordan frame in the following section.

For black holes in Brans-Dicke type theories, the conclusion is the same as in the simple Einstein-scalar theory: the no-hair theorems imply that that the metric is Kerr and the scalar field is constant, so that the scalar charge $q$ vanishes, the mass $M$ is constant in time, and the motion is geodesic.  For bodies ``intermediate'' between weak-field and black holes, such as compact objects like neutron stars, one must solve the field equations and determine the mass and charge via the surface integrals \eqref{MES}-\eqref{qES}.  One interesting example is provided by reference \cite{damour-espositofarese-prl}, which identified a class of Einstein-frame scalar-tensor theories without potential in which compact objects can develop large scalar charges even in the case of small self-gravity.  Their notion of scalar charge, inherited from the general work of \cite{damour-espositofarese}, agrees with our notion up to normalization.  Accounting for a field redefinition relating their scalar field to ours, the precise relationship between our charge $q$ and the charge $\omega$ of \cite{damour-espositofarese-prl} is given by $q=\sqrt{4\pi} \omega$.

While we have performed all analysis for general conformal factor, we note for completeness that in the original Brans-Dicke theory the conformal factor is given by $\Omega^2(\phi) = e^{\frac{\phi}{\sqrt{2\omega+3}}}$.

\subsection{Jordan Frame Brans-Dicke Theory}\label{sec:jordan-frame}

Although the Brans-Dicke theory can be analyzed in the Einstein frame with no loss of generality, for completeness we present the Jordan frame version.  This example also illustrates nicely the generality of the force law, equation \eqref{eomES}, since it applies in both frames, with the meaning of $q$ and $M$ differing in precisely the way that ensures consistency of physical predictions.  The Lagrangian for Jordan frame Brans-Dicke theory is
\begin{equation}
L = \frac{1}{16 \pi} \left( \phi R - \omega g^{ab} \frac{\nabla_a \nabla_b \phi}{\phi} \right),
\end{equation}
where $\omega$ is a constant.  The associated field equation operators are
\begin{align}
16 \pi E^{[g]}_{ab} & = \phi G_{ab} + \Box \phi g_{ab}  - \nabla_a \nabla_b \phi - 8 \pi T^{[\phi]}_{ab} \nonumber \\
16 \pi E^{[\phi]} & = R - \omega \phi^{-2} (\nabla \phi)^2 + 2 \omega \phi^{-1} \Box \phi,
\end{align}
with
\begin{equation}
8 \pi T^{[\phi]}_{ab} = \omega \phi^{-1}\left( \nabla_a \phi \nabla_b \phi - \frac{1}{2} (\nabla \phi)^2 g_{ab} \right),\label{TphiBD}
\end{equation}
where $(\nabla \phi)^2=g^{ab}\nabla_a \phi \nabla_b \phi$ and $\Box \phi=g^{ab} \nabla_a \nabla_b \phi$.  Note that these equations may be combined to yield
\begin{equation}\label{sneaky}
\Box \phi = \frac{16 \pi}{3+2\omega}\left( \phi E^{[\phi]} + g^{ab} E^{[g]}_{ab} \right).
\end{equation}
It is usually assumed that matter does not couple fundamentally to the scalar field, $E^{[\phi]}=-\sigma=0$.  We then have for the non-electrovacuum equations that
\begin{align}
\phi G_{ab} + \Box \phi g_{ab} - \nabla_a \nabla_b \phi & = 8 \pi ( T^{[\phi]}_{ab} +  T_{ab}) \label{nonevgBD} \\
\Box \phi & = \frac{16 \pi}{3+2\omega} g^{ab} T_{ab}. \label{nonevphiBD}
\end{align}

The symplectic potential for the theory is
\begin{align}
16 \pi \Theta^a &= 2 \phi g_{bc} \nabla^{[a}\delta g^{b]c} + \nabla_b \phi \delta g^{bc} \nonumber \\ & - \nabla^a \phi g_{bc} \delta g^{bc} -2 \omega \phi^{-1} \nabla^a \phi \delta \phi,
\end{align}
and the symplectic current is
\begin{align}
16 \pi ( \omega^a[&\{g,\phi\},\{h,\alpha\},\{j,\beta\}] - \phi \omega^a_{GR} ) = -\frac{1}{2} h j^{ab} \nabla_b \phi \nonumber \\ & + j \nabla^a \alpha  + \omega \phi^{-1}\beta (-\nabla^a \phi h + 2h^{ab} \nabla_b\phi + 2 \nabla^a \alpha ) \nonumber \\ & + \beta (\nabla^a h - \nabla_b h^{ab}) - \{h,\alpha\} \leftrightarrow \{j,\beta\},
\end{align}
where, as before, we view $\omega$ as a function of $h^{ab} = \delta_1 g^{ab}$, $j^{ab}=\delta_2 g^{ab}$, $\alpha=\delta_1 \phi$ and $\beta = \delta_2 \phi$, and use the background metric $g_{ab}$ to raise and lower indices on $h$ and $j$.  For the mass $M$ (equation \eqref{Mcoord} or \eqref{Mcov}) we let $h^{\mu \nu} \rightarrow a^{[g]\mu \nu}/r$, $j^{\mu \nu} \rightarrow u^\mu u^\nu=\delta^\mu_{\ 0} \delta^\nu_{\ 0}$, $\alpha \rightarrow a^{[\phi]}/r$ and $\beta \rightarrow 0$.  Keeping only the most singular ($1/r^2$) terms, we have
\begin{align}
16 \pi & \omega^i[\{g,\phi\}, \{\frac{a^{[g]}}{r},\frac{a^{[\phi]}}{r}\}, \{u^\mu u^\nu,0\} ] = \nonumber \\ & \phi \left\{ \partial_k \left( \frac{a^{[g]}_{ki}}{r} \right) - \partial_i \left( \frac{a^{[g]}_{kk}}{r} \right) \right\} - \partial^i \left( \frac{a^{[\phi]}}{r} \right) + O\left( \frac{1}{r} \right).
\end{align}
The mass $M$ is then given by equation \eqref{Mcov} as
\begin{align}
M = \frac{1}{16\pi} \lim_{r \rightarrow 0} \int r^2 d\Omega \ n^i \Bigg[ \phi & \left\{ \partial_k \left( \frac{a^{[g]}_{ki}}{r} \right) - \partial_i \left( \frac{a^{[g]}_{kk}}{r} \right) \right\} \nonumber \\ & - 2 \partial_i \left( \frac{a^{[\phi]}}{r} \right) \Bigg].
\end{align}
Taking into account the appearance of $a^{[g]}_{\mu \nu}$ and $a^{[\phi]}$ in the near-zone background fields, equations \eqref{gbar0} and \eqref{phibar0}, we may rewrite this formula in terms of the near-zone background field configuration,
\begin{align}
M = \frac{1}{16\pi} \lim_{\bar{r}\rightarrow \infty} \int \bar{r}^2 d\Omega \ n^i \Big[ \bar{\phi}^{(0)} & \left( \partial_k \bar{g}^{(0)}_{ki} - \partial_i \bar{g}^{(0)}_{kk} \right) \nonumber \\ & - 2 \partial_i \bar{\phi}^{(0)}  \Big].\label{MBD}
\end{align}
For the scalar charge $q$, equation \eqref{qcov} instructs us to let $h^{\mu \nu} \rightarrow a^{[g]}/r$, $j^{\mu \nu} \rightarrow 0$, $\alpha \rightarrow a^{[\phi]}/r$ and $\beta \rightarrow 1$.  Keeping only the most singular ($1/r^2$) terms, we have
\begin{align}
16 \pi \omega^i[ & \{g,\phi\},\{\frac{a^{[g]}_{\mu \nu}}{r},\frac{a^{[\phi]}}{r}\},\{0,1\} ] = \nonumber \\ & 2 \omega \phi^{-1} \partial^i \left( \frac{a^{[\phi]}}{r} \right) + \partial^i \left( \frac{a^{[g]}_{00}}{r} \right) \nonumber \\ & + \partial_k \left( \frac{a^{[g]}_{ki}}{r} \right) - \partial_i \left( \frac{a^{[g]}_{kk}}{r} \right)  + O\left( \frac{1}{r} \right).
\end{align}
Thus the charge $q$ is given by
\begin{align}
q = - & \frac{1}{16\pi} \lim_{r \rightarrow 0} \int r^2 d\Omega n^i \Bigg[ 2 \omega \phi^{-1} \partial_i \left( \frac{a^{[\phi]}}{r} \right) \nonumber \\ & + \partial_i \left( \frac{a^{[g]}_{00}}{r} \right)  + \partial_k \left( \frac{a^{[g]}_{ki}}{r} \right) - \partial_i \left( \frac{a^{[g]}_{kk}}{r} \right) \Bigg].
\end{align}
Taking into account the appearance of $a^{[g]}_{\mu \nu}$ and $a^{[\phi]}$ in the near-zone background fields, equations \eqref{gbar0} and \eqref{phibar0}, we may rewrite this formula in terms of the near-zone background field configuration,
\begin{align}\
q  = - \frac{1}{16\pi} \lim_{\bar{r} \rightarrow \infty} \int \bar{r}^2 & d\Omega n^i \Bigg[ 2 \frac{\omega}{\bar{\phi}^{(0)}} \partial_i \bar{\phi}^{(0)} + \partial_i \bar{g}^{(0)}_{00} \nonumber \\ & + \partial_k \bar{g}^{(0)}_{ki} - \partial_i \bar{g}^{(0)}_{kk} \Bigg]. \label{qBD}
\end{align}
In the Einstein frame, the mass formula involved only the metric, while the scalar charge formula involved only the scalar field.  In the Jordan frame, both formulae involve both fields.

\subsubsection{Weak-field Bodies}

To put the formulae for the mass and charge to use we must find stationary solutions to the non-electrovacuum equations, \eqref{nonevgBD}-\eqref{nonevphiBD}.  (Since the Brans-Dicke theory is invariant under our scalings, we do not need alternative equations for the near-zone exterior.)  We begin with weak field bodies.  Assuming only linear deviations from a background $\{\eta_{\mu \nu},\hat{\phi}\}$ (where $\hat{\phi}$ is a constant) and taking $T_{00}=\rho$ and $T_{\mu i}=0$, we find (following the original work \cite{brans-dicke}) that a gauge may be chosen where 
\begin{align}
\nabla^2 \alpha & = \frac{- 8 \pi \rho}{3+2\omega} \\
\nabla^2 \gamma_{00} & = - 8\pi \hat{\phi}^{-1} \rho \frac{5+4 \omega}{3+2\omega} \\
\nabla^2 \gamma_{ij} & = - 8 \pi \hat{\phi}^{-1} \rho \frac{1}{3+2\omega} \delta_{ij},
\end{align}
where the fields are given by
\begin{align}
g_{\mu \nu} & = \eta_{\mu \nu} + \gamma_{\mu \nu} - \frac{1}{2} \eta_{\mu \nu} \eta^{\alpha \beta} \gamma_{\alpha \beta} \\
\phi & = \hat{\phi} + \alpha
\end{align}
The solution satisfying our boundary conditions is
\begin{align}
\phi & = \hat{\phi} + \frac{2}{3+2\omega} \frac{M_B}{r} + O\left( \frac{1}{r^2} \right) \\
g_{00} & = -1 + 4 \hat{\phi}^{-1} \frac{2+\omega}{3+2\omega}\frac{M_B}{r} + O\left( \frac{1}{r^2} \right) \\
g_{ij} & = \delta_{ij} \left( 1 + 4 \hat{\phi}^{-1} \frac{1+\omega}{3+2\omega} \frac{M_B}{r} \right) + O\left( \frac{1}{r^2} \right),
\end{align}
where we have defined the total baryonic mass $M_B$,
\begin{equation}
M_B = \int \rho d^3 x.
\end{equation}
Plugging this solution (as $\bar{g}^{(0)},\bar{\phi}^{(0)}$) into equations \eqref{MBD} and \eqref{qBD} gives the mass and charge as
\begin{align}
M & = M_B \\
q & = 0.
\end{align}
This reproduces the fact that weak-field bodies in Brans-Dicke theory behave identically to their counterparts in general relativity: the mass is just the baryonic mass, and the charge vanishes, ensuring geodesic motion and constancy of $M=M_B$.  We remind the reader that $q$ is related to the sensitivity $s$ by equation \eqref{sq}, so that the sensitivity vanishes for weak-field bodies.  Since the sensitivity is dimensionless and vanishes when the mass equals the baryonic mass $M_B$, it is often referred to as the gravitational binding energy per unit mass.

\subsubsection{Black Holes}

For a black hole in Brans-Dicke theory, the metric is given by the Kerr metric and the scalar field is equal to its asymptotic value $\hat{\phi}$.  Adopting asymptotically isotropic coordinates, we have
\begin{align}
g_{00} & = -1 + \frac{2M_{GR}}{r} + O\left( \frac{1}{r^2} \right) \\
g_{ij} & = \delta_{ij} \left( 1 + \frac{2M_{GR}}{r} \right) + O\left( \frac{1}{r^2} \right) \\
\phi & = \hat{\phi},
\end{align}
where $M_{GR}$ is the mass parameter of the black hole, equal to the ADM mass of the metric.  Evaluating the mass and charge (equation \eqref{MBD} and \eqref{qBD}) of the black hole solution gives
\begin{align}
M & = \hat{\phi} M_{GR} \label{MMGR} \\
q & = - \frac{1}{2} M_{GR} \label{qMGR}
\end{align}
The sensitivity of the black hole, equation \eqref{sq}, is given by
\begin{equation}
s=1/2,
\end{equation}
in agreement with previous work.  From the constraint $dM/d\phi=-q$ (equation \eqref{MqES}), it follows that the mass parameter $M_{GR}$ evolves with the external scalar field value $\hat{\phi}$ by
\begin{equation}\label{MGRM0}
M_{GR} = M_0 \hat{\phi}^{-1/2},
\end{equation}
where $M_0$ is some constant independent of $\hat{\phi}$, which might be viewed as the ``intrinsic mass'' of the black hole.  In the weak-field limit,  $\hat{\phi}^{1/2}$ has the interpretation of being proportional to the local value of Newton's constant $\hat{G}$.  Equation \eqref{MGRM0} says that it is the combination $\hat{G} M_{GR}$, rather than the mass parameter $M_{GR}$, that remains constant for a black hole.  Combining equations \eqref{MMGR} and \eqref{MGRM0} gives
\begin{equation}
M = M_0 \sqrt{\hat{\phi}}.
\end{equation}
Thus the mass of the black hole scales as $\hat{G}^{-1/2}$.  If one constructs a ``modified Planck mass'' using $\hat{G}$ rather than $G$, then the black hole mass scales as the Planck mass. This scaling is sometimes assumed a priori, leading to an independent argument that the sensitivity of a black hole is $1/2$.

\subsection{Generalized Jordan Frame Brans-Dicke Theory}
A simple generalization of Brans-Dicke theory studied recently by \cite{alsing-berti-will-zaglauer} is given by the following Lagrangian,
\begin{equation}
L = \frac{1}{16 \pi} \left( \phi R - \omega(\phi) g^{ab} \frac{\nabla_a \nabla_b \phi}{\phi} + M(\phi) \right).
\end{equation}
Here the Brans-Dicke parameter $\omega$ has been promoted to a function of $\phi$, and one has added a potential term denoted $M(\phi)$.  While these changes affect the equations of motion of the theory, they do not alter the symplectic potential or the symplectic current.  Thus, the formulae for the mass and charge are unchanged, and equations \eqref{MES} and \eqref{qES} hold for this theory as well.  However, for a non-zero potential $M(\phi)$, the theory will not be scale-invariant and the construction of the near-zone metric will involve the subtleties discussed in section \ref{sec:nonscaleinv}.  In particular, to construct the near-zone metric one must interpolate between a full solution in the interior and a solution of the equation without the potential term in the asymptotic region.  (Alternatively, a full solution may be used, with equations \eqref{MES} and \eqref{qES} evaluated at an intermediate radius.)  More details of this procedure are given in section \ref{sec:nonscaleinv}.

\section{Vector-Tensor Theories}\label{sec:vector-tensor}

As examples of vector-tensor theories we will consider Einstein-Maxwell theory and the theory of Will and Nordvedt \cite{will-nordtvedt}.  This latter theory offers a simple example of a non-Maxwell vector-tensor theory with second-order field equations.  Many other such theories can be constructed \cite{will-book}; however, the simple Will-Nordvedt theory suffices to illustrate the features of our approach and the general properties of motion in the vector-tensor case.  It is straightforward to repeat these calculations for a more general vector-tensor theory and derive the analogs of equations \eqref{MWN}-\eqref{ehatWN}, below.  

Vector-tensor theories whose vector field is constrained to be unit timelike are known as Einstein-Aether theories \cite{gasperini,jacobson-mattingly,aereview}.  Since a Lagrange multiplier scalar field is used to enforce the unit constraint, these theories are of a scalar-vector-tensor type according to our classification.  However, while the Lagrangian is perfectly covariant, the scalar field does not behave appropriately in the buffer zone\footnote{For example, for a static linearized solution with a point mass source in the Einstein-Aether theory with $c_2=c_4=0, c_1=-c_3$ (in the notation of \cite{aereview}), the Lagrange multiplier field itself is a delta function, instead of having the desired $1/r$ behavior possessed by the other fields.  Our formalism assumes that \textit{all} fields in the Lagrangian have this behavior.} and must be eliminated as a dynamical field before the methods of this paper would apply.  This elimination does not fit naturally into the framework developed in sections \ref{sec:lagrangian}-\ref{sec:motion}, and a discussion of Einstein-Aether theory will be saved for a future publication.

\subsection{Einstein-Maxwell Theory}\label{sec:einstein-maxwell}

The Lagrangian is
\begin{equation}
L = \frac{1}{16\pi} \left( R - g^{ab} g^{bd} F_{ab} F_{cd} \right),
\end{equation}
where $F_{ab} = 2 \nabla_{[a}A_{b]}$.  The field equation operators are
\begin{align}
16\pi E^{[g]}_{ab} & = G_{ab} - 2 \left( F_a{}^c F_{bc} - \tfrac{1}{4} g^{ab} g^{bd} F_{ab} F_{cd} \right) \\
E^{[A]a} & = \tfrac{1}{4\pi} \nabla_b F^{ba}
\end{align}
The symplectic potential is
\begin{equation}
\Theta^a = \Theta^a_{GR} - \frac{1}{4\pi} F^{ab} \delta A_b,
\end{equation}
and the symplectic current is
\begin{align}
\omega^a - \omega^a_{GR} & = 4 \left( h^{ac} F_{cd} \beta^b + F^a{}_d h^{bd} \beta_b + 2 \beta_b \nabla^{[a}\alpha^{b]} \right) \nonumber \\
& \qquad - \{h,\alpha\} \leftrightarrow \{j,\beta\},
\end{align}
where $h^{ab}=\delta_1 g^{ab}$, $j^{ab}=\delta_2 g^{ab}$, $\alpha_a = \delta_1 A_a$, $\beta_a = \delta_2 A_a$, and indices on these quantities are raised and lowered with the background metric.  The Einstein-Maxwell theory has an additional gauge symmetry that may be used to simplify the calculations.  This gauge symmetry gives rise to an additional charge-conservation identity for the theory, $\nabla_a E^{[A]a}=0$.  If this identity is used at the linearized level during the calculations of section \ref{sec:usebianchi}, one finds that $\hat{e}=0$ and $u^a \nabla_a e=0$, i.e., that the hatted charge vanishes and the ordinary charge is conserved.  (These calculations are done explicitly in paper I.)  To provide formulae for the non-vanishing charges we use the symplectic current in equations \eqref{Mcoord}-\eqref{ehatcoord} and rewrite in the near-zone, following the pattern established with previous calculations for scalar-tensor theories.  A short calculation then yields
\begin{align}
M & = \frac{1}{16\pi} \lim_{\bar{r} \rightarrow \infty} \int \bar{r}^2 d\Omega \ n^i \left( \partial_k \bar{g}^{(0)}_{ki} - \partial_i \bar{g}^{(0)}_{kk} \right) \label{MEM} \\
e & = \frac{-1}{4\pi} \lim_{\bar{r} \rightarrow \infty} \int \bar{r}^2 d\Omega \ n^i \partial_i \bar{A}^{(0)}_0 \label{eEM}.
\end{align}
We thus see that the mass $M$ is simply the ADM mass, while the charge $e$ is simply the electric charge as defined by Gauss' law.  Using that $\hat{e}=0$ and  $u^a \nabla_a e=0$, the equations of motion \eqref{eomcov}-\eqref{massevcov} reduce to
\begin{align}
M u^b \nabla_b u^a & = 2 e u_b \nabla^{[b} A^{a]} \\
u^a \nabla_a M & = 0.
\end{align}
That is, the body follows the Lorentz force law and the mass (as well as the charge) is conserved.  We thus recover the standard result for the motion of bodies, while connecting $M$ and $e$ directly to near-zone solutions via the surface-integral formulae \eqref{MEM} and \eqref{eEM}.  We also recover the standard point particle charge-current and stress-energy, as may be seen from equations \eqref{EA1cov}-\eqref{Eg1cov} with $\hat{e}=0$.

\subsection{Will-Nordtvedt Theory}\label{sec:will-nordtvedt}

The Lagrangian for the Will-Nordvedt theory \cite{will-nordtvedt,will-book} is
\begin{equation}\label{LWN}
L = \frac{1}{16\pi} \left( R + \nabla_a A_b \nabla^a A^b \right).
\end{equation}
The field equation operators are
\begin{align}
16 \pi E^{[A]a} & = -2 \nabla_b \nabla^b A^a \label{EAWN} \\
16 \pi E^{[g]}_{ab} & = G_{ab} - \tfrac{1}{2} g_{ab} \nabla_c A_d \nabla^c A^d \nonumber \\ & + \nabla_c \left(  A^c \nabla_{(a} A_{b)} - A_{(a} \nabla^c A_{b)} - A_{(a} \nabla_{b)}
A^c \right) \nonumber \\ & + \nabla_a A^c \nabla_b A_c + \nabla^c A_a \nabla_c A_b, \label{EgEA}
\end{align}
and the symplectic potential is given by
\begin{align}
16\pi (\Theta^a - \Theta^a_{GR})&  = \delta g^{bc} \left( A_c \nabla^a A_b +  A_c \nabla_b A^a - A^a \nabla_b A_c \right) \nonumber \\ & + 2 \delta A_b \nabla^a A^b.
\end{align}
Taking an antisymmetrized variation to obtain the symplectic current, one obtains a long expression.  Since only terms in which a derivative operator acts on a variation can affect the formula for the mass and charges, we only display these terms.  The formula is
\begin{align}
16\pi(\omega^a-\omega_{GR}^a) & = - 2 \kappa^{ab} \beta_b - 2 \kappa^{(ab)} j_{bc} A^c + A^a j_{bc} \kappa^{bc} \nonumber \\ & - \{h,\alpha\}\leftrightarrow\{j,\beta\} \nonumber \\ & + (\textrm{terms with no $\nabla$ on $\{h,\alpha,j,\beta\}$}),
\end{align}
where
\begin{equation}
\kappa_{ab} = \nabla_a \alpha_b + \tfrac{1}{2} A^c(\nabla_a h_{bc} + \nabla_b h_{ac} - \nabla_c h_{ab} ),
\end{equation}
and, as in the Einstein-Maxwell case, we have set $h^{ab}=\delta_1 g^{ab}$, $j^{ab}=\delta_2 g^{ab}$, $\alpha_a = \delta_1 A_a$ and $\beta_a = \delta_2 A_a$, and indices on these quantities are raised and lowered with the background metric.  Furthermore, indices on $\kappa_{ab}$ are raised and lowered with the background metric.  (We note that $\kappa_{ab}$ is just the variation of the derivative of the field, $\kappa_{ab}=\delta_1 (\nabla_a A_b)$.)  Following the pattern of calculations established previously in the paper, we find near-zone formulae for the mass, charge, and hatted charge are,
\begin{align}
M & = \frac{1}{16\pi} \lim_{r \rightarrow \infty} \int r^2 d \Omega \ n^i \Big(  \partial_k g_{ki} - \partial_i g_{kk} \nonumber \\ & \qquad \qquad + 2(A^0 \nabla_0 A_i - A_i \nabla_0 A^0 - A^0 \nabla_i A^0 )\Big) \label{MWN} \\
e & = \frac{-1}{8\pi} \lim_{r \rightarrow \infty} \int r^2 d \Omega \ n_i \nabla^i A^0 \label{eWN} \\
\hat{e} & = \frac{-1}{8\pi} \lim_{r \rightarrow \infty} \frac{1}{A^k A_k}\int r^2 d \Omega \ n_i A_j \nabla^i A^j . \label{ehatWN}
\end{align}
Here we have dropped bars and superscript $(0)$ for convenience; however, as in analogous formulae elsewhere, one should use the near-zone background configuration $\{\bar{g}^{(0)}_{ab},\bar{A}^{(0)a}\}$ in these formulae.   (In particular $\nabla$ is the derivative operator associated with $\bar{g}^{(0)}$.  Note also that the near-zone configuration is stationary; covariant time-derivatives in these formulae could be eliminated in terms of partial spatial derivatives of the metric.) It is understood that when $\hat{A}^i=0$ then the charge $\hat{e}$ vanishes.  (Recall that $\hat{A}^a$ gives the asymptotic value of the vector field, $\lim_{\bar{r} \rightarrow \infty} \bar{A}^{(0)a}=\hat{A}^a$.)  Equations \eqref{MWN}-\eqref{ehatWN} provide formulae for the charges associated with a body in the Will-Nordvedt theory.  Once the field equations have been solved for a given body type, these equations, combined with equation \eqref{eomcov} with $q=0$, may be used to determine the motion of that type of body.  Since the hatted charge is in general non-vanishing and the ordinary charge is not conserved, the Will-Nordvedt theory---and presumably other more general vector-tensor theories---allow a much richer phenomenology of motion than Einstein-Maxwell theory.  This phenomenology remains to be explored.

\vspace{.5cm}

\acknowledgements{I wish to thank Enrico Barausse, Brendan Foster, Fabian Schmidt, Ted Jacobson and Clifford Will for helpful conversations.  This research was supported in part by NASA through the Einstein Fellowship Program, Grant PF1-120082.}

\end{document}